\documentclass[nofootinbib]{revtex4}

\usepackage[percent]{overpic}
\usepackage{natbib}
\usepackage{graphicx}
\usepackage{epsfig}
\usepackage{comment}
\usepackage{amsmath}
\input{epsf}
\usepackage{psfrag}

\usepackage[usenames,dvipsnames]{color}

\newcommand{\beq}{\begin{eqnarray}}
\newcommand{\eeq}{\end{eqnarray}}
\newcommand{\Slash}[1]{{\ooalign{\hfil/\hfil\crcr$#1$}}}

\newcommand{\nn}{\nonumber \\}

\usepackage{amsmath,color,amsfonts}
\usepackage{amssymb,dsfont}
\usepackage{bm}
\usepackage{graphicx}
\usepackage{multirow}

\begin{document}

\title{Quantum entanglement in electron-nucleus collisions: \\Role of the  linearly polarized gluon distribution
}

\author{Michael Fucilla}
\affiliation{ National Centre for Nuclear Research, Pasteura 7, Warsaw 02-093, Poland}

\author{Yoshitaka Hatta  }
\affiliation{Physics Department, Brookhaven National Laboratory, Upton, NY 11973, USA}
\affiliation{RIKEN BNL Research Center, Brookhaven National Laboratory, Upton, NY 11973, USA}

\author{Bo-Wen Xiao}
\affiliation{School of Science and Engineering, The Chinese University of Hong Kong (Shenzhen),
Longgang, Shenzhen, Guangdong, 518172, P.R. China} 
\affiliation{Southern Center for Nuclear-Science Theory (SCNT), Institute of Modern Physics,
Chinese Academy of Sciences, Huizhou, Guangdong 516000, China}

\begin{abstract}
  We calculate the spin density matrix of a back-to-back quark-antiquark pair inclusively produced in electron-nucleus scattering, taking into account the gluon saturation effect and the linearly polarized gluon distribution. We then investigate concurrence and stabilizer Rényi entropy, quantifying entanglement, Bell-nonlocality, and magic. We find that the linearly polarized gluon distribution tends to enhance the entanglement  of a heavy quark pair when the total and relative transverse momenta of the pair are orthogonal. 
\end{abstract}

\maketitle

\section{Introduction}

The interface between quantum information science and high-energy collider physics has recently evolved into a concrete experimental program. The observation of spin correlations in top–antitop ($t\bar t$) production at the LHC by the ATLAS~\cite{ATLAS:2023fsd} and CMS~\cite{CMS:2024pts} collaborations has demonstrated that quantum correlations among quarks can be experimentally accessed. These measurements have stimulated renewed theoretical interest in characterizing entanglement and Bell-nonlocality in QCD processes~\cite{Barr:2024djo,Afik:2022kwm,Afik:2025ejh}. While the top quark provides a particularly clean laboratory at hadron-hadron colliders, because it decays before hadronization, lighter quark systems can offer complementary opportunities in lepton–hadron scattering. \\

In the context of entanglement observables in DIS, earlier work in collinear factorization has investigated spin correlations and Bell-nonlocality in inclusive $q \bar{q}$-pair production at leading order and highlighted the prospects for maximal entanglement in certain kinematic regimes~\cite{Qi:2025onf} (see also \cite{Fucilla:2025kit,Cheng:2025zaw,Xi:2025feb,Hatta:2025obw,Zhang:2026nwm}). In the high-energy limit, however, the appropriate theoretical framework is no longer collinear factorization but $k_T$-factorization, where the transverse momentum of small-$x$ gluons is treated explicitly, and the relevant observables can be expressed in terms of transverse-momentum–dependent gluon distributions. In particular, Deep Inelastic Scattering (DIS) at the future Electron-Ion Collider (EIC) will enable precision studies of quark–antiquark pair production in the small-$x$ regime, where gluon densities become large and nonlinear QCD dynamics sets in. This regime is naturally described within the Color Glass Condensate (CGC) effective theory~\cite{Gelis:2010nm}, which provides a systematic description of the phenomena of gluon saturation. This latter constitutes one of the central pillars of the physics program of the Electron–Ion Collider. From this perspective, the investigation of $q \bar q$ spin-correlations in a CGC framework paves the way for a novel research direction uniting the physics of dense gluonic matter with the quantum-informational characterization of spin entanglement\footnote{For previous works on different types of entanglement in the context of high-energy QCD, see~\cite{Kovner:2015hga,Peschanski:2016hgk,Kharzeev:2017qzs,Liu:2018gae,Ramos:2020kyc,Bhattacharya:2024sno,Guo:2024jch,Brandenburg:2024ksp,Hatta:2024lbw,Dumitru:2025bib,Bloss:2025ywh,Agrawal:2025yoe,Ouchen:2025tta,Hentschinski:2025pyq,Golec-Biernat:2025hwa,Zhang:2025ean}}. In a recent work, we analyzed spin–spin entanglement in exclusive diffractive $q\bar q$ production, where the interaction proceeds via color-singlet exchange~\cite{Fucilla:2025kit,Hatta:2025obw}. At high-energy, where the color-singlet exchange is described in terms of the celebrated hard (BFKL) Pomeron of QCD~\cite{BFKL1,BFKL2,BFKL3}, it was observed that the $q\bar q$ diffractively produced always exhibits entanglement and Bell-nonlocality~\cite{Fucilla:2025kit}. In the context of the collinear factorization framework based on Generalized Parton Distributions (GPDs), the same spin correlations were investigated in~\cite{Hatta:2025obw}, showing a more complex pattern of entanglement between the quark and the antiquark. \\

The main aim of the present paper is the natural continuation of this program, i.e., the computation of spin correlations in inclusive $q \bar{q}$ production in electron-nucleus collisions in the small-$x$ factorization framework. Compared to the exclusive (diffractive) channel, the inclusive process benefits from significantly larger event rates, which enhances its experimental viability. At present, access to spin correlations in heavy-quark production relies primarily on the fragmentation of heavy quarks into heavy baryons, followed by their semi-leptonic decays. This strategy is intrinsically limited by small branching fractions and reduced reconstruction efficiencies.\footnote{See, however, Ref.~\cite{Cheng:2025cuv} for an alternative method to measure spin correlations in the light-quark sector.}
The inclusive channel at the EIC and the LHC (in Ultra Peripheral Collisions, UPCs) offers a substantially more favorable environment, both in terms of luminosity and statistical precision, making it a particularly promising avenue for future studies of spin entanglement in DIS. In this work, we will consider the production of the $q\bar{q}$-pair in the back-to-back limit. We will be particularly interested in how the linearly polarized gluon distribution \cite{Mulders:2000sh,Boer:2010zf} affects the spin density matrix, and consequently, the degrees of entanglement, Bell-nonlocality, and magic.\\

\section{The spin-density matrix for the inclusive dijet production at small-$x$}
\begin{figure}
    \centering
    \includegraphics[width=0.3\linewidth]{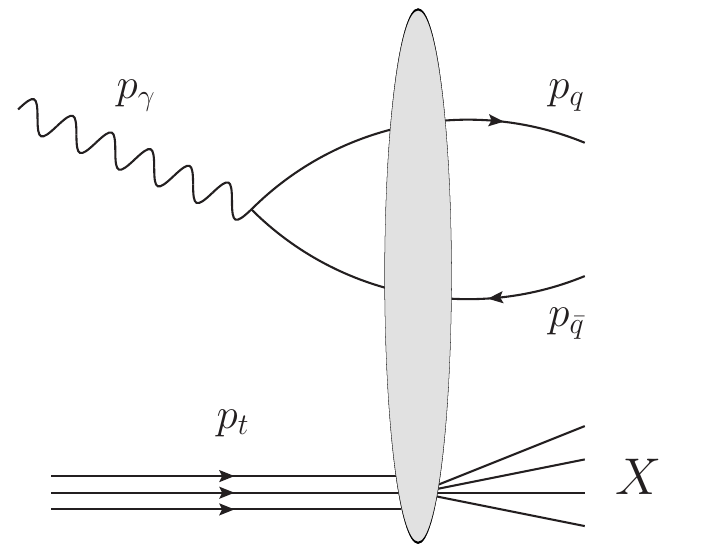}
    \caption{Schematic representation of the inclusive dijet production in DIS or photoproduction at small-$x$ in the dipole picture: a high-energy photon fluctuates into a quark–antiquark dipole, which then scatters eikonally off the dense gluonic target. There is no rapidity gap between the $q \bar{q}$-pair and the undetected system $X$.}
    \label{InclusiveDijet}
\end{figure}
The total cross-section of the inclusive dijet production in DIS or photoproduction has long been considered in the small-$x$ literature~\cite{Dominguez:2011wm}, including both next-to-leading logarithmic~\cite{Caucal:2021ent,Taels:2022tza,Caucal:2023fsf,Bergabo:2023wed,Iancu:2022gpw,Caucal:2024nsb} and next-to-eikonal corrections~\cite{Altinoluk:2022jkk,Altinoluk:2024zom,Kovchegov:2025gcg,Mukherjee:2026cte,Chirilli:2026vij}. In the dipole picture, it can be visualized as in Fig.~\ref{InclusiveDijet}. Here, a fast-moving photon of momenta $p_{\gamma}$ in the $x^3$-direction fluctuates into a quark-antiquark pair of momenta $p_q$ and $p_{\bar{q}}$, respectively. The projectile dipole then interacts with the dense gluonic target (either a proton or nuclei), moving ultrarelativistically in the opposite direction with momenta $p_t$. At very high energies, the interaction is eikonal, i.e., it occurs with a small deflection angle. Introducing the standard Sudakov decomposition $k^{\mu} = k_1^+ n_1^{\mu} + k_2^+ n_2^{\mu} + k_T^{\mu}$, where $n_1^{\mu}$ and $n_2^{\mu}$ specify the +/- directions and such that $n_1 \cdot n_2 = 1$, we will can parametrize the 4-momenta as
\begin{gather}
    p_{\gamma}^{\mu} = p_{\gamma}^{+} n_1^{\mu} - \frac{Q^2}{2 p_{\gamma}^+} n_2^{\mu} \; ,  \hspace{2 cm} p_{t}^{\mu} = p_{t}^{-} n_2^{\mu} \; , \nonumber \\   p_q^{\mu} 
    = z p_{\gamma}^+ n_1^{\mu} + \frac{-p_{qT}^{2} + m^2 }{2 z p_{\gamma}^+} n_2^{\mu} + p_{q T}^{\mu} \; ,  \qquad p_{\bar{q}}^{\mu} = \bar{z} p_{\gamma}^+ n_1^{\mu} + \frac{-p_{\bar{q}T}^{2} + m^2 }{2 \bar{z} p_{\gamma}^+} n_2^{\mu} + p_{\bar{q} T}^{\mu} \; ,
\end{gather}
where $m$ is the current quark mass, and we introduced the longitudinal momentum fractions $z$ and $\bar{z}\equiv 1-z$. The target mass has been neglected. 
We denote with $p_{\perp}^i$ the Eucledian counterpart of the Minkowskian 4-vector $p_T^{\mu}$, i.e. $p_{\perp}^2 = - p_{T}^2$. \\

In the following, we study this process within the framework of $k_T$-factorization, where the cross section at small-$x$ is dominated by the BFKL Pomeron. Furthermore, we rely on the semi-classical small-$x$ effective field theory, commonly referred to as the Color Glass Condensate (CGC). In this approach, the fast partons in the target are treated as static classical color sources generating a strong background field, while the slow gluons are described as dynamical gauge fields propagating in this background. The interaction of the projectile dipole with the target is encoded in Wilson lines,
\begin{equation}
\mathcal{U} ( x_\perp) = \mathcal{P} \exp\left[ i g_s \int dx^+  T_R \cdot A^-(x^+, x_\perp) \right],
\end{equation}
which resum multiple eikonal scatterings of the quark and antiquark off the target color field. Observables such as the inclusive dijet cross section can then be expressed in terms of correlators of Wilson lines, whose small-$x$ evolution is governed by the corresponding nonlinear evolution equations.

\subsection{General structure of the spin-density matrix}

The cross section for inclusive dijet production in the CGC framework is given by \cite{Dominguez:2011wm} (see also Appendix \ref{App:DijetCross} for more details)

\begin{gather}
     \frac{d \sigma^{L/T}_{\alpha\alpha'\beta\beta'} }{d z d^2 p_{q \perp} d^2 p_{\bar{q} \perp} } =   N_c \alpha_{em} e_f^2 \; p_{\gamma}^+ \int \frac{d^2 x_{1 \perp}}{(2 \pi)^2}  \int \frac{ d^2 x_{2 \perp}}{(2 \pi)^2} \int \frac{d^2 x_{1'  \perp}}{(2 \pi)^2}  \int \frac{d^2 x_{2' \perp}}{(2 \pi)^2} e^{-i p_{q \perp} \cdot x_{1 1' \perp} } e^{-i p_{ \bar{q} \perp} \cdot x_{2 2' \perp} } \nonumber \\  \times \psi^{L/T (\lambda) *}_{\beta\beta'} (x_{1'2'\perp}) \psi^{L/T (\lambda)}_{\alpha\alpha'} (x_{12\perp})    \bigg( 1 + S^{(4)} (x_{1 \perp}, x_{2 \perp}; x_{2' \perp}, x_{1' \perp}) - S^{(2)} (x_{1 \perp}, x_{2 \perp}) -  S^{(2)} (x_{2' \perp}, x_{1' \perp}) \bigg) \; ,
     \label{start}
\end{gather}
where the dipole and quadrupole matrix elements are
\beq
S^{(2)} (x_{1 \perp}, x_{2 \perp}) = \frac{1}{N_c} \langle {\rm Tr_c} [\mathcal{U}_F (x_{1 \perp}) \mathcal{U}_F^{\dagger} (x_{2 \perp}) ]  \rangle_{A} \; ,
\eeq
\beq
S^{(4)} (x_{1 \perp}, x_{2 \perp}; x_{2' \perp}, x_{1' \perp} ) = \frac{1}{N_c} \langle {\rm Tr_c} [\mathcal{U}_F (x_{1 \perp}) \mathcal{U}_F^{\dagger} (x_{2 \perp}) \mathcal{U}_F (x_{2' \perp}) \mathcal{U}_F^{\dagger} (x_{1' \perp})]  \rangle_{A} \; .
\eeq
$\alpha$ and $\alpha'$ are the spinor indices of the quark and antiquark in the amplitude, respectively, and $\beta$ and $\beta'$ are those for the complex-conjugate amplitude. The unpolarized cross section is obtained by setting $\alpha=\alpha'$ and $\beta=\beta'$ and summing over these indices, but we keep them open for the purpose of computing the spin density matrix.
The $\gamma \to q\bar{q}$ splitting wave functions $\psi^{L/T}$ are well known \cite{Nikolaev:1990ja,Brodsky:1994kf}, see below. 

In order to perform the coordinate integrations in (\ref{start}), it is convenient to introduce  the dijet relative and total transverse momenta as 
\beq
P_\perp = \bar{z}p_{q\perp}-zp_{\bar{q}\perp}, \qquad q_\perp = p_{q\perp}+p_{\bar{q}\perp}, 
\eeq
and their conjugate coordinates 
\beq
u_\perp = x_{1 \perp} - x_{2 \perp} \; , \hspace{0.5 cm} u'_\perp = x_{1'\perp} - x_{2'\perp} \; , \hspace{0.5 cm} v_\perp = z x_{1 \perp} + \bar{z} x_{2 \perp} \; , \hspace{0.5 cm} v_\perp' = z x_{1' \perp } + \bar{z} x_{2' \perp}. 
\eeq
This transformation has a unit  Jacobian 
\begin{gather}
\int d^2x_{1 \perp} d^2x_{2 \perp} d^2 x_{1' \perp} d^2 x_{2' \perp}  e^{-i p_{q \perp} (x_1-x_{1'} )_\perp - i p_{\bar{q} \perp} (x_2-x_{2'})_\perp} \; ... \nonumber \\ = \int d^2 u_\perp d^2 u^\prime_\perp d^2v_\perp d^2 v_\perp^\prime  e^{-iP_\perp (u-u^\prime)_\perp -iq_\perp (v-v^\prime)_\perp} \; ... \,
\end{gather} 
Naively, by Fourier transforming in $u_\perp$ and $u'_\perp$, one finds the product $\psi^*(P_\perp)\psi(P_\perp)$, where  
\beq
\psi^{L}_{\alpha\alpha'} (P_{\perp}) &=& \int d^2 x_{12\perp} e^{-i x_{12 \perp} \cdot P_{\perp}} \psi^{L/T(\lambda)}_{\alpha\alpha'} (x_{12\perp}) 
\nn 
&=& (2 \pi)^2  \sqrt{\frac{1}{p_q^+ p_{\bar{q}}^+ p_{\gamma}^+ }  }  \frac{z \bar{z} Q}{P_\perp^2+z\bar{z}Q^2+m^2} \bar{u}_{\alpha} (p_q)\gamma^+ v_{\alpha'} (p_{\bar{q}}) \; ,
\label{Eq:PsiLon}
\eeq
\beq
\psi^{T(\lambda)}_{\alpha\alpha'} (P_\perp) &=& (2 \pi)^2  \sqrt{\frac{1}{8 p_q^+ p_{\bar{q}}^+ p_{\gamma}^+ }} \frac{\varepsilon^i_\lambda}{P_\perp^2+z\bar{z}Q^2+m^2} \bar{u}_{\alpha} (p_q) \left((1-2z)P_\perp^i+i\epsilon^{ij}P_\perp^j \gamma_5-m\gamma_\perp^i\right) \gamma^+ v_{\alpha'} (p_{\bar{q}}) \: , \nonumber \\
\label{Eq:PsiTrans}
\eeq
representing a $q\bar{q}$ pair with relative momentum $P_\perp$. However, this simple interpretation is spoiled by the fact that the dipole and quadrupole amplitudes depend nontrivially on $u_\perp,u'_\perp$. In general, one finds unequal momenta $\psi^*(k_\perp)\psi(k'_\perp)$ in the amplitude and the complex-conjugate amplitude due to finite transverse momentum transfer from the target. In order to avoid this complication,  in the present paper, we consider the correlation limit, corresponding to a kinematical configuration in which the two jets are back-to-back. More precisely,  the individual transverse momenta of the two jets are much larger than their sum $|P_{\perp}| \simeq |p_{q \perp}| \simeq |p_{\bar{q} \perp}| \gg |p_{q \perp} + p_{\bar{q} \perp}| \simeq |q_{\perp}| $. In coordinate space, this corresponds to the configuration $u_\perp, u_\perp' \ll v_\perp,v_\perp'$. We can then   expand the target matrix elements around $u_\perp, u_\perp' \simeq 0$ and find \cite{Dominguez:2011wm}
\beq
 && \left[ 1 + S^{(4)} (x_{1 \perp}, x_{2 \perp}; x_{2' \perp}, x_{1' \perp}) - S^{(2)} (x_{1 \perp}, x_{2 \perp}) -  S^{(2)} (x_{2' \perp}, x_{1' \perp}) \right] \nonumber \\  && = - u_{\perp i} u^\prime_{\perp j} \frac{1}{N_c} \langle \text{Tr}_{\rm c} [(\partial_i \mathcal{U} (v_\perp)) \mathcal{U}^\dagger(v_\perp^\prime) (\partial_j \mathcal{U} (v_\perp^\prime)) \mathcal{U}^\dagger(v_\perp)] \rangle_A \equiv u_{\perp i} u^\prime_{\perp j}  G_{ij}(v_\perp,v'_\perp).
\eeq
$G_{ij}$ can be identified with the Weisz\"acker-Williams (WW) gluon 
 distribution in the small-$x$ limit 
\begin{gather}
G_{ij} (v_\perp,v'_\perp)
= \frac{g_s^2}{N_c} \int_{-\infty}^{\infty} d v^+ d v'^+  \langle \text{Tr}_{\rm c} [F^{i-} ( v_{\perp} ) \mathcal{U}^{[+] \dagger} F^{j-} ( v_{\perp}' ) \mathcal{U}^{[+]} ] \rangle 
\; ,
\end{gather} 
where $\mathcal{U}^{[+]}$ is the half-infinite staple-shaped Wilson line extending to future  infinity $x^+=\infty$.  
The factor $u_\perp$ can be traded for the derivative with respect to $P_\perp$ acting on the photon wavefunction, leading to 
\beq
\frac{d\sigma^{L/T}_{\alpha\alpha'\beta\beta'} }{dz d^2P_\perp d^2q_\perp} &=& \frac{N_c \alpha_{em} e_f^2 \; p_{\gamma}^+}{(2 \pi)^8} \int d^2u d^2u' e^{-iP_\perp (u-u')_\perp } u_{\perp i} u'_{\perp j} \psi^{L/T(\lambda) *}_{\beta\beta'}(u_\perp')  \psi_{\alpha\alpha'}^{L/T(\lambda)} (u_\perp) \nonumber \\ && \times \int d^2v_\perp d^2v'_\perp G_{ij}(v_\perp,v'_\perp) e^{-iq_\perp( v-v')_{\perp}}
\nn  
&=& \frac{N_c \alpha_{em} e_f^2 \; p_{\gamma}^+}{{(2 \pi)^8}} \frac{\partial}{\partial  P_{\perp}^{'i} } \frac{\partial}{\partial P_{\perp }^j }  \psi^{L/T(\lambda) *}_{\beta\beta'}(P'_\perp) \psi^{L/T(\lambda)}_{\alpha\alpha'} (P_\perp) \Big\arrowvert_{P'_\perp=P_\perp} \nonumber \\ && \times  \left[\frac{\delta_{ij}}{2}G_0(q_\perp)+\left(\frac{q_\perp^i q_\perp^j}{q_\perp^2}-\frac{\delta_{ij}}{2}\right) G_2(q_\perp)\right] \; ,  \label{ll}
\eeq
where we have decomposed the WW gluon TMD $G_{ij}(q_\perp)$ into the unpolarized part $G_0$ and the linearly polarized part $G_2$ \cite{Mulders:2000sh}. 
In deriving (\ref{ll}), we have observed that the Fourier transform and also the derivative in $P_\perp$ do not act on  the spinors $\bar{u}(p_q)$ and $v(p_{\bar{q}})$. This can be justified in the shockwave (dipole) frame in which the quark and the antiquark are moving fast in the same direction. Technically, it is because only the `good' components of the spinors contribute \cite{Beuf:2016wdz}. If one works in generic frames, in particular in the center-of-mass (CM) frame of the $q\bar{q}$ pair, the $P_\perp$-dependence of the spinors must be fully taken into account.

 We emphasize that the factorization of the $P_\perp$ and $q_\perp$ dependencies in (\ref{ll}) occurs only in the correlation limit. This is an essential simplification for the purpose of computing the spin density matrix. Away from this limit, the quark and the antiquark do not have fixed momenta common in the amplitude and the complex-conjugate amplitude. One then has to deal with  entanglement not only in  spin space but also in  momentum space. 
 
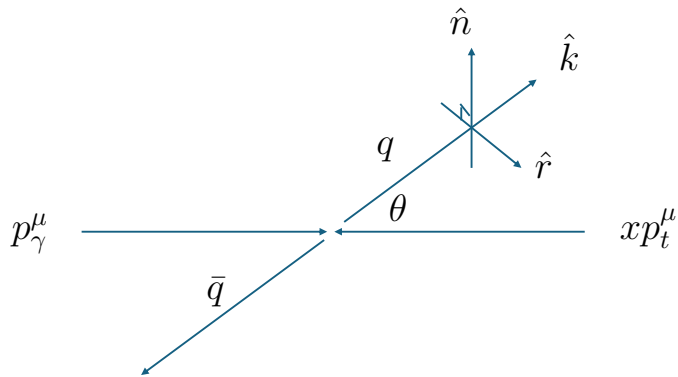
\begin{figure}[t]
        \begin{overpic}[width=\textwidth]{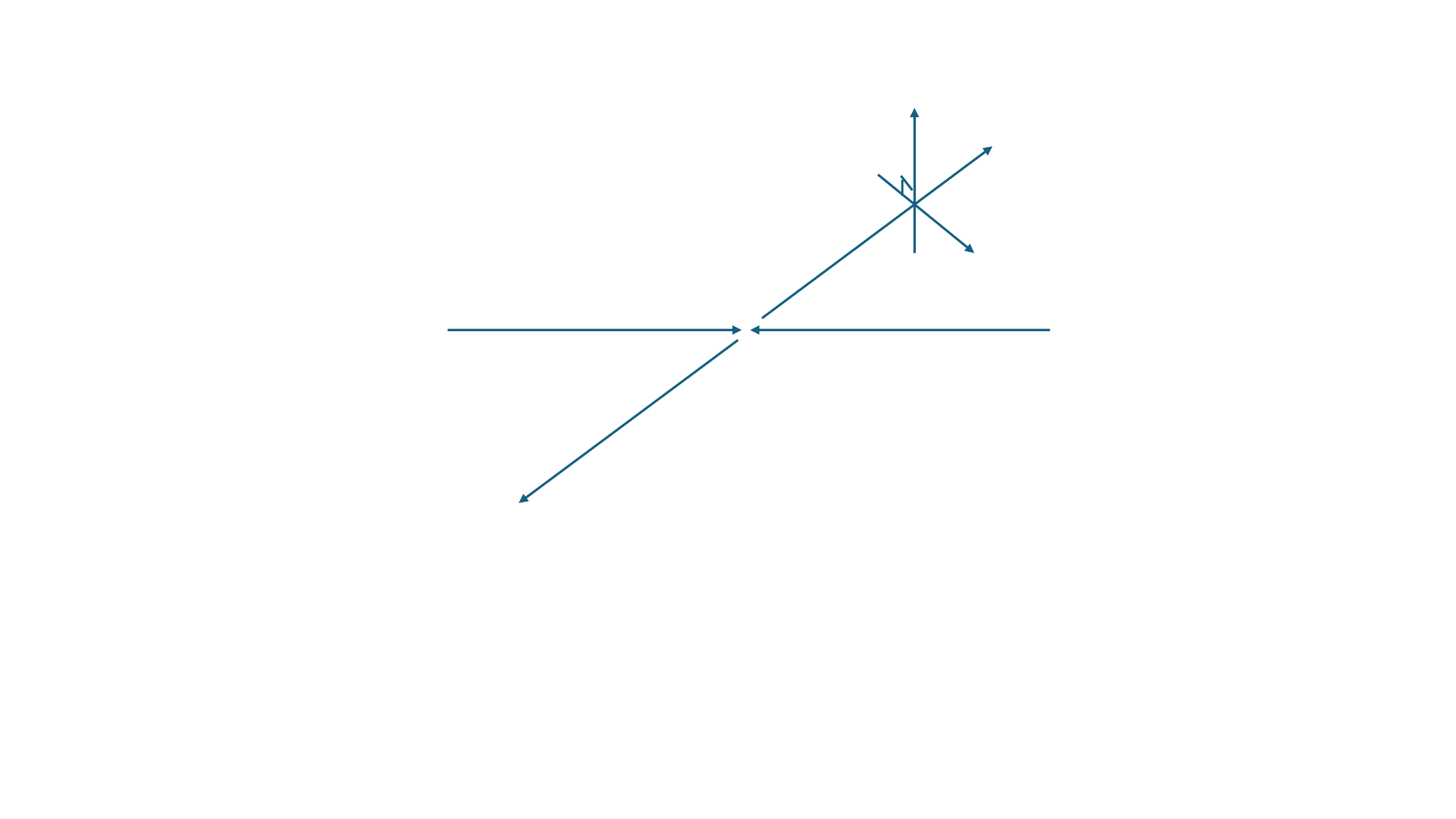}
            \put(56,34.5){\Large $\theta$ }
            \put(61,50){\Large $\hat{n}$ }
            \put(68,38){{\Large $\hat{r}$}} 
            \put(25,33){{\Large $p_\gamma^\mu$}}
             \put(75,33){{\Large $xp_t^\mu$}} 
           \put(70,47){{\Large $\hat{k}$}}
           \put(55,40){{\Large $q$}}
            \put(41,28){{\Large $\bar{q}$}}  
        \end{overpic} \vspace{-40mm}
    \caption{Center-of-mass (CM) frame of the quark-antiquark pair at $q_\perp\approx 0$. $x=\frac{Q^2+M^2}{2p_\gamma\cdot p_t}$ is the momentum fraction of the target entering the hard scattering.  }
         \label{pic}
  \end{figure}

We shall evaluate the hard factor in the following form 
\beq
&& \left[ \frac{\partial}{\partial  P_{\perp}^i } \psi_{\beta\beta'}^{L/T(\lambda)*} (P_\perp) \right]  \left[ \frac{\partial}{\partial P_{\perp }^j } \psi^{L/T(\lambda)}_{\alpha\alpha'} (P_\perp) \right] 
=\xi_\alpha^\dagger \eta_{\alpha'}^\dagger \label{Eq:GeneralHardFactor}\\ && \times \left[\delta^{ij}\left(A^{(0)}_{L/T}\mathbb{I}\otimes \mathbb{I}  +   (\tilde{ C}^{(0)}_{L/T})_{ab} \sigma^a \otimes \sigma^b \right)   +\left(\frac{2P_\perp^i P_\perp^j}{P_\perp^2}-\delta^{ij}\right)\left(A^{(2)}_{L/T} \mathbb{I}\otimes \mathbb{I}  +   (\tilde{C}^{(2)}_{L/T})_{ab} \sigma^a \otimes \sigma^b \right)\right]\xi_\beta \eta_{\beta'} \; , \notag
\eeq
where $\mathbb{I}$ is the 2x2 unit matrix and $\sigma^{a}$ are the Pauli matrices. These matrices are defined in a special frame  in which the quark and the antiquark have back-to-back momenta, see Fig.~\ref{pic}
\beq
\tilde{p}_q^\mu =(\tilde{p}_q^0,\tilde{p}_q^n,\tilde{p}_q^r,\tilde{p}_q^k)= \frac{M}{2}(1,0,0,\beta), \qquad \tilde{p}_{\bar{q}}^\mu = \frac{M}{2}(1,0,0,-\beta),
\eeq
where 
\beq
M=\sqrt{\frac{P_\perp^2+m^2}{z\bar{z}}}, \qquad \beta = \sqrt{1-\frac{4m^2}{M^2}}, \qquad \cos \theta = \frac{2z-1}{\beta}. 
\label{Eq:KinematicVar}
\eeq 
$\theta$ is the scattering angle in the pair's CM frame. 
This frame can be reached from the original `shockwave frame' as follows. First, we perform the $P_\perp$-derivatives in (\ref{Eq:GeneralHardFactor}). As discussed above, the external spinors  $\bar{u}(p_q),v(p_{\bar{q}})$ do not depend on $P_\perp$, see (\ref{delphi}). We then boost along the $x^3$-direction (assuming $q_\perp \approx 0$) to reach the CM frame of the pair. Finally, we perform a 3-dimensional rotation by angle $\theta$ of the coordinate system $(\hat{x}^1,\hat{x}^2,\hat{x}^3) \to (\hat{n},\hat{r},\hat{k})$ such that $\hat{k}$ coincides with  the direction of the quark and $\hat{n}$ is normal to the reaction plane. $\xi_\alpha$ and $\eta_\beta$ with $\alpha,\beta=\pm$ are the two-component spinors for the quark and the antiquark, respectively, with the sign  $\pm$ referring  to the projection of their spins along the $\hat{k}$ direction (i.e., the same as helicity for the quark and opposite to helicity for the antiquark.) They act on the first and second Pauli matrices $\sigma^{a=n,r,k}$, namely, $\xi^\dagger \eta^\dagger (\sigma^a\otimes \sigma^b) \xi\eta \equiv \xi^\dagger \sigma^a \xi \eta^\dagger \sigma^b \eta$.

Inserting (\ref{Eq:GeneralHardFactor}) into (\ref{ll}),  we obtain  
\beq
\frac{d\sigma^{L/T}_{\alpha\alpha'\beta\beta'} }{d z d^2 P_\perp d^2 q_\perp} 
&=& \frac{N_c \alpha_{em} e_f^2 \; p_{\gamma}^+}{(2 \pi)^8} \xi^\dagger_\alpha\eta^\dagger_{\alpha'} \Biggl\{ \left(G_0A_{L/T}^{(0)}+\cos (2 \phi_{P,q})  G_2A^{(2)}_{L/T}\right)\mathbb{I}\otimes \mathbb{I} \nn && \qquad +\left(G_0(\tilde{C}^{(0)}_{L/T})_{ab}+ \cos (2 \phi_{P,q}) G_2(\tilde{C}^{(2)}_{L/T})_{ab}\right) \sigma^a\otimes \sigma^b \Biggr\} \xi_\beta \eta_{\beta'} , \notag
\eeq 
where $\phi_{P,q} = \phi_P-\phi_q$ is the relative angle between $\phi_P$ and $\phi_q$. 
The physical cross section/density matrix is a linear combination of the longitudinal and transverse contributions
\beq
\frac{d\sigma_{\alpha\alpha'\beta\beta'} }{d z d^2 P_\perp d^2 q_\perp} = 
  \frac{N_c \alpha_{em} e_f^2 \; p_{\gamma}^+}{(2 \pi)^8} {\cal A}\xi^\dagger_\alpha \eta^\dagger_{\alpha'}\left(\mathbb{I}\otimes \mathbb{I}+{\cal C}_{ab}\sigma^a\otimes \sigma^b\right)\xi_\beta \eta_{\beta'}
   \label{Eq:General_Spin_Density_Matrix}
\eeq
  where 
  \beq
  {\cal A}&=&  G_0\left(A^{(0)}_T+\varepsilon A^{(0)}_L\right)+\cos (2 \phi_{P,q})  G_2\left(A^{(2)}_T +\varepsilon A^{(2)}_L\right), \nn 
  {\cal C}_{ab}&=&\frac{1}{{\cal A}}\left[G_0\left(\tilde{C}^{(0)}_{T}+\varepsilon \tilde{C}^{(0)}_{L}\right) + \cos (2 \phi_{P,q}) G_2\left(\tilde{C}^{(2)}_{T}+\varepsilon \tilde{C}_{L}^{(2)}\right)
  \right]_{ab},
  \label{Eq:GenCoeffSpinDen}
  \eeq
  and $\varepsilon$ is the ratio of the transverse and longitudinal photon fluxes ($\varepsilon=0$ in UPC and $\varepsilon\approx 1$ at EIC). By calculating the coefficients $C^{(0)}$ and $C^{(2)}$ for arbitrary photon's virtuality and polarization, we are able to consider both dijet photoproduction ($Q=0$) and dijet electroproduction $(Q\neq 0)$.

If we consider the angular-averaged cross section
\beq
 \int_0^{2 \pi} d \phi_{P,q} \frac{ d\sigma^{L/T} }{d z d^2P_\perp d^2 q_\perp }  \sim  \Bigl[ \mathbb{I}\otimes \mathbb{I} + {\cal C}^{(0)}_{ab}\sigma^a \otimes \sigma^b \Bigr],   \qquad {\cal C}_{ab}^{(0)} &=& \frac{\left(\tilde{C}^{(0)}_T+\varepsilon \tilde{C}^{(0)}_{L}\right)_{ab}}{ A^{(0)}_T+\varepsilon A^{(0)}_L } \;  , \label{g2}
\eeq 
the density matrix does not depend on $G_0$ and $G_2$, namely, it does not depend on the structure of the target.  
By staying differential in the relative angle $\phi_{P,q}$, we can study how the density matrix is influenced by the gluon distributions of the target.

\subsection{Longitudinal photon}
We start by considering the case in which the incoming gluon is longitudinally polarized. From (\ref{Eq:PsiLon}), we  find 
\beq
\left[ \frac{\partial}{\partial  P_{\perp}^i } \psi_{\beta\beta'}^{L*} (P_\perp) \right]  \left[ \frac{\partial}{\partial P_{\perp }^j } \psi^{L}_{\alpha\alpha'} (P_\perp) \right] =   \frac{ (2 \pi)^4}{p_q^+ p_{\bar{q}}^+ p_{\gamma}^+   }  \frac{4 z^2 \bar{z}^2 Q^2 P_{\perp}^i P_{\perp}^j }{ (P_\perp^2 + z \bar{z} Q^2 + m^2)^4} \bar{v}_{\beta'} (p_q) \gamma^+ u_{\beta} (p_{\bar{q}})  \bar{u}_{\alpha} (p_q)\gamma^+ v_{\alpha'} (p_{\bar{q}}) \; . 
\label{Eq:LongProdSplittWave}
\eeq
As we noted below  (\ref{Eq:PsiTrans}), the $P_\perp$-derivative does not act on the spinors. We immediately notice that, in the notation of (\ref{Eq:GeneralHardFactor}),
\beq
A^{(0)}_L=A^{(2)}_L \qquad \tilde{C}^{(0)}_L=\tilde{C}^{(2)}_L \; .
\eeq 
Summing over the helicity indices, we reproduce the known unpolarized cross section 
\beq
A_L=  \frac{4(2\pi)^4z^2\bar{z}^2Q^2P_\perp^2}{p_\gamma^+(P_\perp^2+z\bar{z}Q^2+m^2)^4},
\eeq 
\beq
\frac{d\sigma^{L} }{d z d^2 P_\perp d^2 q_\perp} &=&    \frac{ 16 N_c \alpha_{em} e_f^2 z^2 \bar{z}^2 Q^2 P_\perp^2 }{ (2 \pi)^4(P_\perp^2 + z \bar{z} Q^2 + m^2)^4}  \left(G_0(q_\perp)+\cos(2\phi_{P,q}) G_2(q_\perp)\right). 
\eeq

The spinor product with open helicity indices in (\ref{Eq:LongProdSplittWave}) can be evaluated by following the same procedure as in~\cite{Fucilla:2025kit}.  The result is 
\beq
\bar{v}_{\beta'} (p_q) \gamma^+ u_{\beta} (p_{\bar{q}})  \bar{u}_{\alpha} (p_q)\gamma^+ v_{\alpha'} (p_{\bar{q}}) = \frac{ 8 z\bar{z}  (p_{\gamma}^+)^2}{4} \xi^\dagger_\alpha \eta^\dagger_{\alpha'}\left(\delta_{\alpha\beta}\delta_{\alpha'\beta'}+C^L_{ab}\xi^\dagger_\alpha \sigma^a \xi_\beta \eta^\dagger_{\alpha'}\sigma^b \eta_{\beta'}\right) \xi_\beta\eta_{\beta'}
\eeq
with 

\beq
C_{nn}^L=1, \qquad C_{rr}^L=-C_{kk}^L=-\frac{1-(2-\beta^2)\cos^2\theta}{1-\beta^2\cos^2\theta} = -\frac{P_\perp^2-(1-2z)^2m^2}{P_\perp^2+(1-2z)^2m^2} \; ,
\nonumber \\ C_{rk}^L=C_{kr}^L = -\frac{\sqrt{1-\beta^2}\sin 2\theta}{1-\beta^2\cos^2\theta} = \frac{2(1-2z)\sqrt{P^2_\perp} m}{P_\perp^2+(1-2z)^2m^2} \; ,
\label{Eq:LonSpinDensMatrix}
\eeq
The velocity $\beta$ and the angle in the CM frame are  as defined in (\ref{Eq:KinematicVar}).   We immediately notice that  the $C$-matrix (\ref{Eq:LonSpinDensMatrix}) is independent of the target, and is actually  the same  as in \cite{Qi:2025onf,Fucilla:2025kit,Hatta:2025obw}. In other words, for the longitudinally polarized virtual photon,  the same spin density matrix appears in (i)  one-gluon  exchange \cite{Qi:2025onf}, (ii) two-gluon and quark-antiquark color-singlet exchange \cite{Hatta:2025obw}, (iii) multiple gluon color-singlet (`Pomeron') exchange \cite{Fucilla:2025kit},  and (iv) multiple gluon exchange in inclusive production (this work).  The reason behind this `universality' is interesting to explore. 
As observed in \cite{Qi:2025onf}, the states represented by the density matrix (\ref{Eq:LonSpinDensMatrix}) are maximally entangled pure states with the corresponding concurrence $\mathcal{C}[\rho] \equiv 1$. In the relativistic limit $\beta \rightarrow 1$, or  for the symmetric configuration $z=1/ 2$ (meaning $\cos \theta = 0$ in the CM frame), or along the line $z=\frac{1}{2}\pm \frac{P_\perp}{2m}$ \cite{Hatta:2025obw}, it reduces to one of the Bell states. \\

We now turn our attention to another quantum information theory quantity, the so-called “magic”. This latter quantifies the extent to which a quantum state departs from the class of stabilizer states, which can be efficiently simulated on a classical computer. In high-energy physics, this concept has recently been introduced as a tool to characterize the intrinsic quantum complexity of states produced in scattering processes. Magic can be quantified through the stabilizer Rényi entropy \cite{Leone:2021rzd,White:2024nuc} (see appendix \ref{QuantumInfo}), which, in this case, can be written as a function of the ratio $y=m/|P_{\perp}|$ \cite{Hatta:2025obw},
\begin{gather}
    S_{\rm R}^{\rm stab} (y,z)= - \ln \left(\frac{1 +14 y^4 (1-2 z)^4+ y^8 (1-2 z)^8}{\left( 1 + y^2 (1-2 z)^2\right)^4}\right) \; .
\end{gather}
This function is minimal (indeed zero) at $z = 1/2$, as expected for a Bell state, independently of the value of $y$. For a massless quark, the magic is always zero. For a massive quark, the behavior of the magic is governed by the ratio of its mass to $P_{\perp}$. In general, at fixed $y$, $M^{L}(z)$ is a function that vanishes at $\pm\infty$ and exhibits four local maxima and three local minima. Since the physical domain is restricted to $0<z<1$, for relatively small values of the ratio $y$ ($y \leq 0.4$) only the local minimum at $z = 1/2$ lies within the physical range, and the global maximum is attained at the endpoints ($z = 0,1$). For larger values, $0.4<y<1$, two of the local maxima also fall within the physical domain, where they constitute the absolute maxima.

\subsection{Transverse photon}
\label{sec:TransvPhoton}
The transverse photon case is much more involved. The details of the calculation are provided in  Appendix \ref{App:DetailTransvPhoton}, while here, we limit ourselves to reporting the result. First, we discuss the spin-density matrix elements proportional to the total cross sections. We have
\begin{gather}
    A^{(0)}_T = \frac{2(2\pi)^4}{p_{\gamma}^+} \frac{(z^2+\bar{z}^2)( P_{\perp}^4 + \bar{Q}^4)+2 P_{\perp}^2 m^2}{( P_{\perp}^2 + \bar{Q}^2 )^4} \; ,
\end{gather}
and 
\begin{gather}
    A^{(2)}_T = \frac{4(2\pi)^4}{  p_{\gamma}^+} \frac{\; P_{\perp}^2  \left(m^2 -  \left( z^2 + \bar{z}^2 \right) (m^2 + z \bar{z} Q^2)\right)}{\left(  P_{\perp}^2+\bar{Q}^2\right)^4} \; , 
\end{gather}
where $\bar{Q}^2\equiv z\bar{z}Q^2+m^2$. 
Then, the unpolarized  cross section summed over quark helicities  is given by
\begin{gather}
\frac{d \sigma^{T} }{d z d^2 P_\perp d^2 q_\perp} = \frac{N_c \alpha_{em} e_f^2 \; p_{\gamma}^+}{(2 \pi)^8} \left(A^{(0)}_T G_0 + A^{(2)}_TG_2\cos (2 \phi_{P,q}) \right) 
\; .
\end{gather}

As for the spin density matrix, our result for the angular independent part $C_{T ab}^{(0)} = \tilde{C}_{T ab}^{(0)} /A_T^{(0)}$ is   
\begin{gather}
   C_{T nn}^{(0)} = -\frac{2 z \bar{z} \left( P_{\perp}^4 + \bar{Q}^4 \right)}{2 m^2 P_{\perp}^2+\left( z^2 + \bar{z}^2 \right) \left( P_{\perp}^4 + \bar{Q}^4 \right)} \; ,
\end{gather}

\begin{gather}
    C_{T rr}^{(0)} = \frac{-2 z \bar{z} }{\left(m^2+P_{\perp}^2\right) \left(m^2 (1-2
   z)^2+P_{\perp}^2\right) \left( (z^2 + \bar{z}^2) \left(\bar{Q}^4+P_{\perp}^4\right)+2 m^2   
   P_{\perp}^2\right)}   \nonumber \\ \times   \bigg \{ m^4 \left((1-2 z)^2 \bar{Q}^4 - (3-2 z) (1 + 2 z) P_{\perp}^4\right) \nonumber \\ +2 m^2
   P_{\perp}^2 \left(2 \bar{Q}^2 P_{\perp}^2+ (z^2 + \bar{z}^2) \bar{Q}^4+ (z^2 + \bar{z}^2 - 2)
   P_{\perp}^4\right)-P_{\perp}^4
   \left(\bar{Q}^4+P_{\perp}^4\right) \bigg \} \; ,
\end{gather}

\begin{gather}
    C_{T kk}^{(0)} = \frac{1}{\left(m^2+P_{\perp}^2\right) \left(m^2 (1-2 z)^2+P_{\perp}^2\right) \left( (z^2 + \bar{z}^2) \left(\bar{Q}^4+P_{\perp}^4\right)+2 m^2 P_{\perp}^2\right)} \nonumber \\ \times \bigg \{ -m^4 (1-2 z)^2 \left(4 \bar{Q}^2 P_{\perp}^2 + (z^2 + \bar{z}^2) \bar{Q}^4 - ( 3 + 2 \bar{z} z)
   P_{\perp}^4 \right) \nonumber \\ + 2 m^2 P_{\perp}^2 \left(-2 (z^2 + \bar{z}^2) \bar{Q}^2 P_{\perp}^2 - 4
   \bar{z}^2 z^2 \bar{Q}^4+\left(1-4 \left(z^4-2 z^3+z\right)\right) P_{\perp}^4\right) \nonumber \\ + (z^2 + \bar{z}^2)
   P_{\perp}^4 \left(\bar{Q}^4+P_{\perp}^4\right)+2 m^6 (1-2 z)^2
   P_{\perp}^2 \bigg \} \; ,
\end{gather}

\begin{gather}
  C_{T rk}^{(0)} =  \frac{2 m (1-z) |P_{\perp}| }{\left(m^2+P_{\perp}^2\right) \left(m^2 (1-2
   z)^2+P_{\perp}^2\right) \left( (z^2 + \bar{z}^2) \left(\bar{Q}^4+P_{\perp}^4\right)+2 m^2
   P_{\perp}^2\right)} \nonumber \\ \times  \bigg \{ -m^4 (1-2 z) \left( (1-2 z) \bar{Q}^2 + ( 1 + 2 z ) P_{\perp}^2\right)  + m^2
   \left(-4 z^2 \bar{Q}^2 P_{\perp}^2+(1-2 z)^2 \bar{Q}^4 + \left(8 z^2-4 z-1\right)
   P_{\perp}^4\right) \nonumber \\ + \left(4 z^2-2 z+1\right) \bar{Q}^4 P_{\perp}^2+\bar{Q}^2 P_{\perp}^4 - 2
   z (1-2 z) P_{\perp}^6 \bigg \} \; ,
\end{gather}

\begin{gather}
  C_{T kr}^{(0)} =  \frac{ - 2 z m  |P_{\perp}| }{\left(m^2+P_{\perp}^2\right)
   \left(m^2 (1-2 z)^2+P_{\perp}^2\right) \left( (z^2 + \bar{z}^2) \left(\bar{Q}^4+P_{\perp}^4\right)+2
   m^2 P_{\perp}^2\right)} \nonumber \\ \times \bigg \{ m^4 (2 z-1) \left((1-2 z) \bar{Q}^2+(2 z-3) P_{\perp}^2\right)+m^2
   \left(-4 (1-z)^2 \bar{Q}^2 P_{\perp}^2+(1-2 z)^2 \bar{Q}^4 + (3 - 4 z (3- 2 z) )
   P_{\perp}^4\right) \nonumber \\ + \left(4 z^2-6 z+3\right) \bar{Q}^4 P_{\perp}^2+\bar{Q}^2
   P_{\perp}^4+\left(4 z^2-6 z+2\right) P_{\perp}^6 \bigg \} \; ,
\end{gather}
and all other vanishing. Remarkably, the matrix $C^{(0)}_{T}$ exactly agrees with that of the one-gluon exchange approximation $\gamma^*+g\to q+\bar{q}$ calculated in the collinear factorization framework~\cite{Qi:2025onf}.  Our derivation based on  the photon light-cone wavefunction demonstrates the robustness of this matrix against higher order corrections (multiple gluon exchanges), and at the same time provides a highly  nontrivial check of the result~\cite{Qi:2025onf}. However, unlike in the longitudinal case, $C^{(0)}_T$ does not agree with the corresponding matrix in the color-singlet (Pomeron) exchange \cite{Fucilla:2025kit,Hatta:2025obw}.   
In the massless case, the matrix $C_T^{(0)}$ drastically simplifies and becomes diagonal   
\begin{gather}
C_T^{(0)}= \begin{pmatrix} -\frac{2 z \bar{z} }{ z^2 + \bar{z}^2  } & 0 & 0 \\ 
0 & \frac{2 z \bar{z} }{ z^2 + \bar{z}^2  } & 0\\
0 & 0 & 1 \end{pmatrix}. \label{masslessc0}
\end{gather}
Note that $C^{(0)}_{Tnn}=-C^{(0)}_{Trr}$. This result leads to the corresponding concurrence $\mathcal{C}[\rho] = \tfrac{2 z \bar{z} }{ z^2 + \bar{z}^2  }$. We find that the mass correction to the concurrence is always negative.

For the linearly polarized part, the remaining normalized  coefficients $C_{T ab}^{(2)} \equiv \tilde{C}_{T ab}^{(2)} /A_T^{(2)}$ are\footnote{To avoid confusion, it should be noted that $C^{(2)}_T$ itself is not a physical spin correlation matrix. In practice we only need  $\tilde{C}^{(2)}_T=A_T^{(2)}C^{(2)}_T$ when computing the physical density matrix (\ref{Eq:GenCoeffSpinDen}).}
\begin{gather}
   C_{T nn}^{(2)} = \frac{2 \left(m^2+ z \bar{z} Q^2 \right)}{2 m^2 - Q^2 \left(z^2+ \bar{z}^2 \right)} \; ,
\end{gather}

\begin{gather}
    C_{T rr}^{(2)} = \frac{2 z \bar{z} \left(m^2 \left((1-2 z)^2 \bar{Q}^2+2 P_{\perp}^2\right)-\bar{Q}^2
   P_{\perp}^2\right)}{\left( m^2 - ( z^2 + \bar{z}^2 ) \bar{Q}^2 \right) \left(m^2 (1-2
   z)^2+P_{\perp}^2\right)} \; ,
\end{gather}

\begin{gather}
    C_{T kk}^{(2)} = \frac{ m^2 (1-2  z)^2 \left(  P_{\perp}^2 -(1 + 2 z \bar{z} ) \bar{Q}^2 \right) - (z^2 + \bar{z}^2 ) \bar{Q}^2
   P_{\perp}^2+m^4 (1-2 z)^2}{\left( m^2 - ( z^2 + \bar{z}^2 ) \bar{Q}^2 \right) \left(m^2 (1-2
   z)^2+P_{\perp}^2\right)} \; ,
\end{gather}

\begin{gather}
    C_{T rk}^{(2)} = -\frac{m (1-z) \left(\bar{Q}^2 \left(m^2 (1-2 z)^2+\left(8 z^2-4 z+1\right) P_{\perp}^2\right)+m^2 \left(1-4
   z^2\right) P_{\perp}^2+P_{\perp}^4\right)}{ |P_{\perp}| \left( m^2 - ( z^2 + \bar{z}^2 ) \bar{Q}^2 \right) \left(m^2 (1-2 z)^2+P_{\perp}^2\right)} \; ,
\end{gather}

\begin{gather}
    C_{T kr }^{(2)} = \frac{m z \left(\bar{Q}^2 \left(m^2 (1-2 z)^2+(4 z (2 z-3)+5) P_{\perp}^2\right) - m^2 (3 + 4 z (z-2) )
   P_{\perp}^2+P_{\perp}^4\right)}{ |P_{\perp}| \left( m^2 - ( z^2 + \bar{z}^2 ) \bar{Q}^2 \right) 
   \left(m^2 (1-2 z)^2+P_{\perp}^2\right)}\; ,
\end{gather}
and all other vanishing. These results are new. In the massless limit, one can check that  $C_T^{(2)}=C_T^{(0)}$ as given by (\ref{masslessc0}), and therefore (cf., (\ref{Eq:GenCoeffSpinDen}))  
\beq
{\cal C}_T \equiv   \frac{G_0\tilde{C}_T^{(0)}+\cos(2\phi_{P,q})G_2\tilde{C}^{(2)}_T}{G_0A_T^{(0)}+\cos(2\phi_{P,q})G_2A_T^{(2)}} = C_T^{(0)}. \label{q=0}
\eeq
In the massless limit, after taking the linearly polarized gluon distribution into account, the transverse spin density matrix (without the longitudinal part) gives the same concurrence $\mathcal{C}[\rho] = \tfrac{2 z \bar{z} }{ z^2 + \bar{z}^2  }$. Once we turn on the quark mass, we again find that the concurrence decreases irrespective of $\phi_{P,q}$ and the ratio $G_2/G_0$ (as long as $|G_2/G_0|\leq 1$, as required by the positivity bound). Note that Eq.~(\ref{q=0}) no longer holds once we include the longitudinal contribution with $Q\neq 0$.

Moreover, 
in the massless limit, the matrix $[C_T^{(0)}]^T C_T^{(0)}$  has only two independent eigenvalues
\begin{equation}
    \mu_1 = 1 \;, \hspace{1.5 cm} \mu_2 = \frac{4 (1-z)^2 z^2}{[z^2 + (1-z)^2]^2} \; ,
\end{equation}
whose sum is greater than one. This immediately implies a violation of the Bell-CHSH inequality and an entangled state. When $z=1/2$, we have $\mu_1 + \mu_2 = 2$, implying a maximal violation of the Bell-CHSH inequality, and thus a maximally entangled state. The magic is vanishing at the endpoints ($z=0,1$) and in $z=1/2$, while it reaches its maximum value
\begin{gather}
    S_{\rm R, max}^{\rm stab} \simeq 0.188 \; ,
\end{gather}
for $z \simeq 0.267$ and $z \simeq 0.732$.

\section{Numerical analysis}
\label{Numerical_analysis}

In the case of light quarks, the spin density matrices (longitudinal and transverse) are very simple, and we analyzed their main features in the previous section. Also, the impact of the linearly polarized gluon distribution is minor. Therefore, in the following, we will consider the case of massive quark pair production, which is considerably more complex. To characterize the main features of the density matrix, we will study concurrence, Bell-nonlocality and stabilizer Rényi entropy. Their basic definitions are reviewed in Appendix~\ref{QuantumInfo}.

\begin{figure}
    \centering
\includegraphics[width=0.68\linewidth]{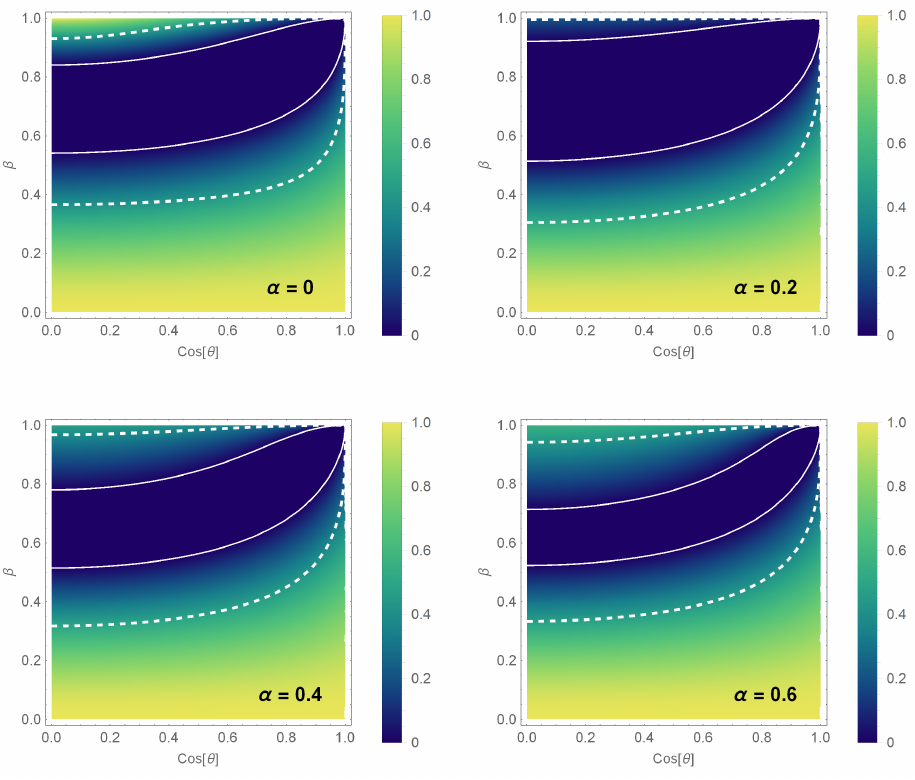}
    \caption{Density plot of the concurrence $\mathcal{C}[\rho]$ associated to the spin density matrix in eq.~(\ref{g2}) with $\varepsilon=1$, as functions of quark velocity $\beta$ and $\cos \theta$, at given values of the virtuality parameter $\alpha=Q^2/M^2$. The value of the concurrence is identified by the color according to the vertical bar, and the continuous and dashed white lines indicate the boundaries for entanglement and Bell-nonlocality, respectively. Namely, the $q\bar{q}$ pair is not entangled between the two solid curves and Bell's inequality cannot be  violated between the two dashed curves.}
    \label{fig:Concurrence_AngularAve_TL}
\end{figure}

\subsection{$\phi_{P,q}$-averaged case}

First consider the $\phi_{P,q}$-averaged spin-density matrix (\ref{g2}). 
In this case,  the target function $G_0$ decouples and  our result is fully equivalent to the one obtained in  collinear factorization~\cite{Qi:2025onf}. We take this opportunity to complement and extend the analysis of Ref.~\cite{Qi:2025onf}. 
In order to facilitate the comparison, we  rewrite the spin-density matrix elements in terms of the three variables $\cos\theta$, $\beta$, and $\alpha \equiv  Q^2/M^2$ (cf., eqs.~(\ref{Eq:KinematicVar}) and~(\ref{Eq:LonSpinDensMatrix})), and study how these elements vary with respect to these variables.

\begin{figure}
    \centering
\includegraphics[width=0.68\linewidth]{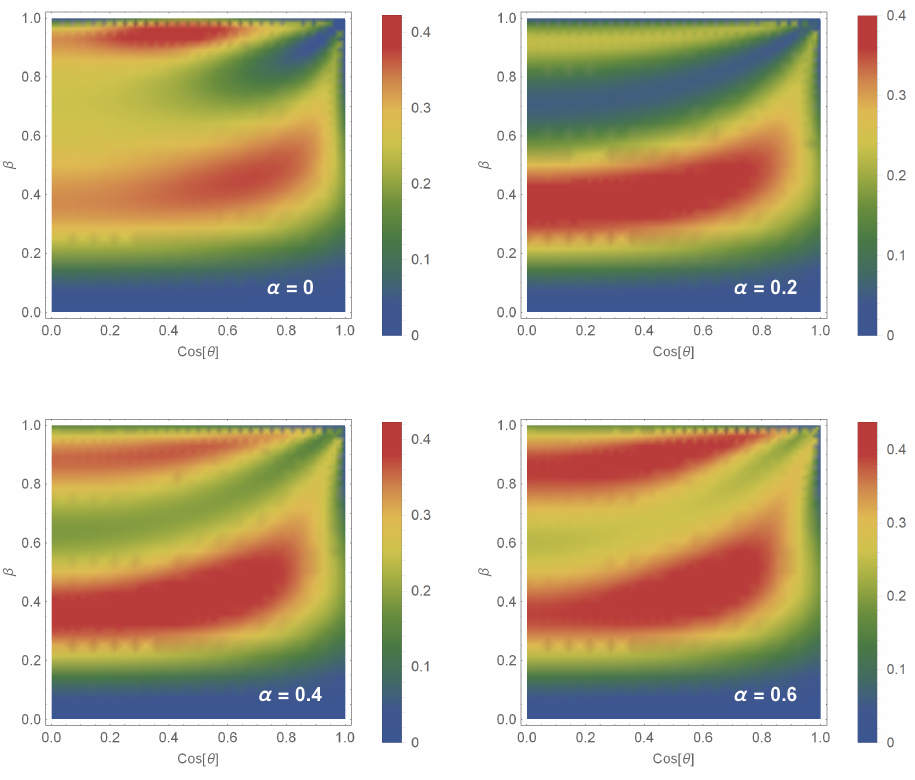}
    \caption{Density plot of the stabilizer R\'enyi entropy $S_{\rm R}^{\rm stab}$ associated to the spin density matrix in eq.~(\ref{g2}) with $\varepsilon=1$, as functions of quark velocity $\beta$ and $\cos \theta$, at given values of the virtuality parameter $\alpha=Q^2/M^2$. The value of $S_{\rm R}^{\rm stab}$ is identified by the color according to the vertical bar.}
    \label{fig:Magic_AngularAve_TL}
\end{figure}

In Ref.~\cite{Qi:2025onf}, the transverse and longitudinal photon cases were studied separately. 
However, in DIS at $Q \neq 0$, the measured cross section is the linear combination of the transverse and longitudinal contributions. Unfortunately, it is difficult to experimentally separate the two contributions at the EIC since typically $\varepsilon \sim 1$ at high energy colliders (see, however, \cite{Klest:2025yik}). In Fig.~\ref{fig:Concurrence_AngularAve_TL}, we thus study the complete spin-density matrix of heavy-quark production in DIS setting $\varepsilon= 1$. The pair is not entangled (separable) in the region between the two solid curves. Bell's inequality cannot be  violated in the region between the two dashed curves. (Outside, the pair exhibits Bell-nonlocality, see Appendix~\ref{QuantumInfo}.) 
From Fig.~3 of \cite{Qi:2025onf}, we see that  the region where the pair is entangled is reduced compared to the case of the transversely polarized photon alone. This is interesting because the pair is always entangled in the purely   longitudinal case  \cite{Qi:2025onf}. Naively, adding  the longitudinal  component makes entanglement  stronger, but this turns out to be not the case. Nonetheless, the entangled region remains substantial and is always larger than the region where Bell-nonlocality is seen. This is in accordance with the general expectation (Bell-nonlocality) $\subset$ (entanglement).

In Fig.~\ref{fig:Magic_AngularAve_TL}, we investigate the magic of the produced system through the stabilizer R\'enyi entropy. The results exhibit a characteristic pattern.  The entropy is suppressed when the pair is either not entangled (i.e., between the solid curves in Fig.~\ref{fig:PhiDepConc1}), or strongly  entangled (i.e.,  $\beta\to 1$ or $\beta\to 0$).  
 Instead, it is nonvanishing in intermediate regions between weak and strong entanglement. Such a tendency has been observed in other reactions \cite{White:2024nuc,Liu:2025qfl,Hatta:2025obw,Gargalionis:2026onv}. 
The dependence on $Q^2$ appears to be non-monotonic, though regions of enhanced  entropy tend to widen as $Q^2$ increases. The largest value of $S^{\rm stab}_{\rm R}$ that we find in this study is about $ 0.45$. In comparison,  the theoretical upper limit for a two-qubit system is $S_{\rm R}^{\rm stab} \le \ln (16 / 7) \simeq 0.827$ \cite{Liu:2025frx}, which is reduced to~\cite{Hatta:2025obw}
\begin{equation}
    S_{\rm R}^{\rm stab} \le \ln \frac{9}{5} \simeq 0.588 \; ,
\end{equation}
in systems with vanishing polarization (i.e., if $B_a=\bar{B}_b=0$ in eq.~(\ref{Eq:GenSpinDens2Qubit})).

\subsection{Dependence of entanglement and magic on $\phi_{P,q}$}

The fully $\phi_{P,q}$-dependent spin-density matrix of heavy-quark production in DIS is given by eq.~(\ref{Eq:General_Spin_Density_Matrix}). In this case, our analytical result is novel and depends on both the angle $\phi_{P,q}$ and the imbalance $q_{\perp}$. Furthermore, as in the diffractive case of Refs.~\cite{Fucilla:2025kit,Hatta:2025obw},  the coefficients ${\cal C}_{ab}$, depend on the gluon distributions of the target proton/nucleus. In particular, at high energy, the result can be sensitive to the small-$x$ dynamics of the target, including the gluon saturation effect. Leaving detailed phenomenological  investigations for future work, here we focus on understanding the general features of the results, in particular whether entanglement and magic exhibit anisotropy with respect to the angle~$\phi_{P,q}$. 

We employ  the McLerran-Venugopalan model ~\cite{McLerran:1993ni} for the WW gluon distribution $G_0$ and the linearly polarized gluon distribution $G_2$~\cite{Metz:2011wb,Dominguez:2011br,Dumitru:2016jku}. 
\begin{gather}
  G_0 (q_\perp)  =  8 \pi S_{\perp} \frac{C_F}{N_c} \int_0^{\infty} d |r_{\perp}| \frac{ J_0 (|q_{\perp }| | r_{\perp} |)  }{|r_{\perp}| }  \left[1-\exp \left(-\frac{Q_{s }^2 |r_{\perp}|^2}{4}   \ln \frac{1}{|r_{\perp}|^2 \Lambda_{\rm IR}^2} \right)\right] \; ,
  \label{Eq:WWUnPo}
\end{gather}
and
\begin{gather}
  G_2(q_\perp)  =  8 \pi S_{\perp} \frac{C_F}{N_c} \int_0^{\infty} d |r_{\perp}| \frac{ J_2 (|q_{\perp }| | r_{\perp} |)  }{|r_{\perp}| \ln \frac{1}{|r_{\perp}|^2 \Lambda_{\rm IR}^2} }  \left[1-\exp \left(-\frac{Q_{s }^2 |r_{\perp}|^2}{4}   \ln \frac{1}{|r_{\perp}|^2 \Lambda_{\rm IR}^2} \right)\right] \; ,
  \label{Eq:WWLinPo}
\end{gather}
where $Q_{s}$ is the saturation momentum. 
The spin-density matrix elements $\mathcal{C}_{ab}$ clearly depend only on the ratio  $G_2(q_\perp)/G_0(q_\perp)$ (see eq.~(\ref{Eq:General_Spin_Density_Matrix})). We will follow the implementation of Ref.~\cite{Dumitru:2016jku}, taking $Q_{s}^2 \simeq 1$ GeV$^2$. Since $G_2(q_\perp)/G_0(q_\perp)\propto q_\perp^2$ as $q_\perp\to 0$,  to see the impact of $G_2$,  one typically needs $q_\perp > Q_{s}$. On the other hand, the present calculation assumes that the pairs are back-to-back  
\beq
P_\perp^2=m^2\beta^2\frac{1-\cos^2\theta}{1-\beta^2} \gg q_\perp^2. \label{cond}
\eeq 
When $q_\perp$ is a few GeV, and for the charm quark $m\sim 1.5$ GeV, the condition (\ref{cond}) is  satisfied only when $\beta\approx 1$, i.e., the relativistic (massless) limit.  However, for the bottom quark $m\sim 4.2$ GeV, we find a reasonable  phase space, say,  $0.6 <\beta < 1$ with a decent window in $\cos\theta$.  

\begin{figure}
    \centering
    \includegraphics[width=0.68\linewidth]{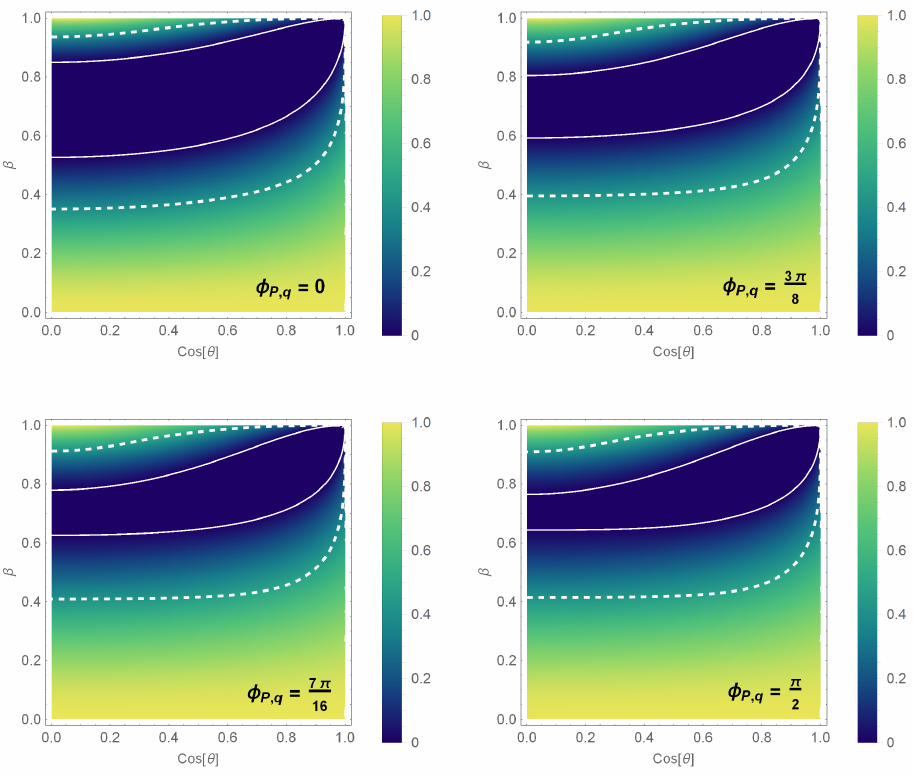}
    \caption{Density plot of the concurrence $\mathcal{C}[\rho]$ associated to the spin density matrix in eq.~(\ref{Eq:General_Spin_Density_Matrix}) for $\alpha=0$ ($Q^2=0$) and $q_{\perp} = 2.5$ GeV, as functions of quark velocity $\beta$ and  $\cos \theta$. The value of the concurrence is identified by the color according to the vertical bar, and the continuous and dashed white lines indicate the boundaries for entanglement and Bell-nonlocality, respectively. }
    \label{fig:PhiDepConc1}
\end{figure}

\begin{figure}
    \centering
    \includegraphics[width=0.68\linewidth]{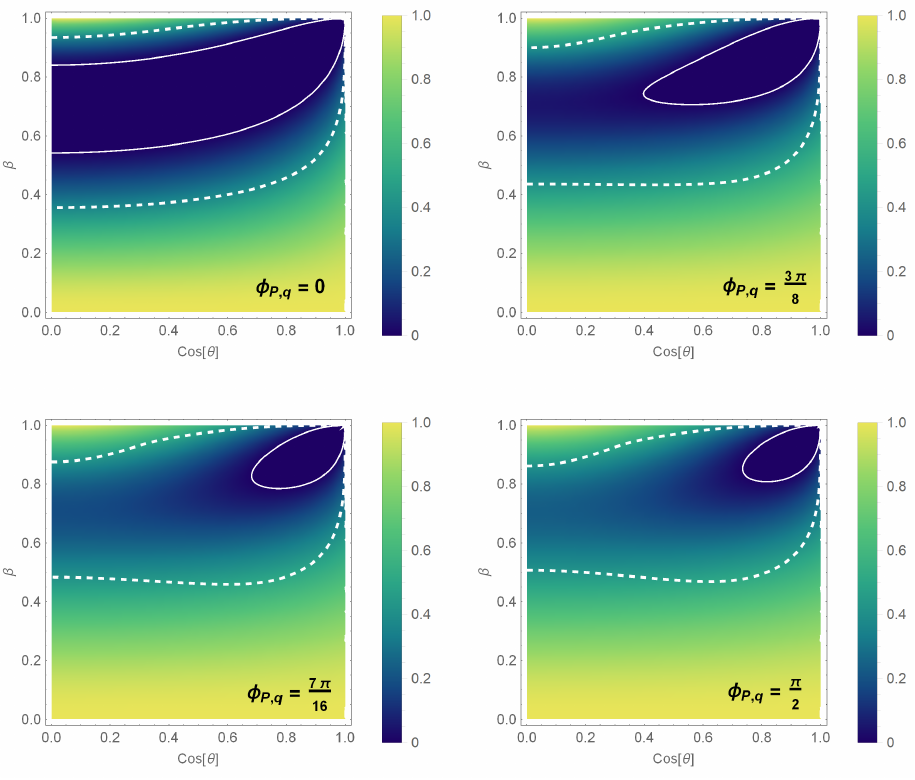}
    \caption{Density plot of the concurrence $\mathcal{C}[\rho]$ associated with the spin density matrix in eq.~(\ref{Eq:General_Spin_Density_Matrix}) for $\alpha=0$ ($Q^2=0$) and $q_{\perp} = 4$ GeV, as functions of quark velocity $\beta$ and the scalar projection $\cos \theta$. The value of the concurrence is identified by the color according to the vertical bar, and the continuous and dashed white lines indicate the boundaries for entanglement and Bell-nonlocality, respectively.}
    \label{fig:PhiDepConc3}
\end{figure}

The results for concurrence and Bell-nonlocality are plotted in Figs.~\ref{fig:PhiDepConc1} 
and~\ref{fig:PhiDepConc3} for $q_\perp=2.5$ GeV and $q_\perp=4$ GeV, respectively, and for different values of $\phi_{P,q}$. (Note that the effect of the  linearly polarized gluon  vanishes at $\phi_{P,q}=\frac{\pi}{4}$). Although the plots cover the entire  phase space in $(\cos\theta,\beta)$ assuming arbitrary  values of $m$, in practice we have in mind  bottom quark pairs, and only the region  $0.6\lesssim\beta$ and $\cos\theta$ not too close to unity is physically relevant. Fortunately, interesting effects take place within this reduced phase space.   We have found that when $\cos 2\phi_{P,q}$ is positive, meaning $0 \le \phi_{P,q} \le \frac{\pi}{4}$ and $\frac{3\pi}{4}\le \phi_{P,q}\le \pi$, there is almost no dependence on $\phi_{P,q}$.  Concurrence is actually reduced, but only by a very small amount of order $10^{-2}$. However, as soon as $\cos 2\phi_{P,q}$ becomes negative, namely when $\frac{\pi}{4}<\phi_{P,q}< \frac{3\pi}{4}$,  we observe that  the region bounded by the solid curve noticeably shrinks, meaning that  $G_2$ tends to enhance the concurrence/entanglement. The effect is more pronounced at higher $q_{\perp}$ because, at a fixed value of the saturation scale $Q_s$, the ratio $G_2 (q_{\perp})/G_0(q_{\perp})$ monotonically increases as $q_{\perp}$ increases. The maximal effect is observed when $P_\perp$ and $q_\perp$ make a right angle $\phi_{P,q}=\frac{\pi}{2}$, as  seen most clearly in the lower  plots of  Fig.~\ref{fig:PhiDepConc3}. In contrast, the boundary of Bell-nonlocality is only weakly affected by the presence of $G_2$.

\begin{figure}
    \centering
    \includegraphics[width=0.68\linewidth]{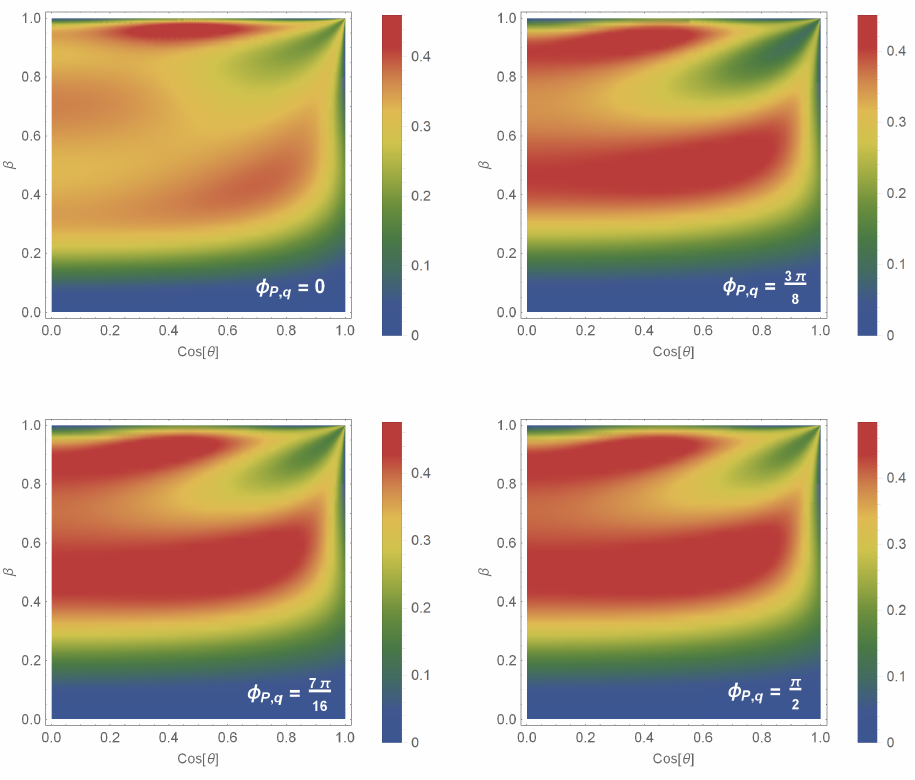}
    \caption{Density plot of the stabilizer R\'enyi entropy $S_{\rm R}^{\rm stab}$ associated to the spin density matrix in eq.~(\ref{Eq:General_Spin_Density_Matrix}) for $\alpha=0$ ($Q^2=0$) and $q_{\perp}=4$ GeV, as functions of quark velocity $\beta$ and the scalar projection $\cos \theta$. The value of $S_{\rm R}^{\rm stab}$ is identified by the color according to the vertical bar.}
    \label{fig:PhiDepMag}
\end{figure}

From Fig.~\ref{fig:PhiDepMag}, we observe that the R\'enyi entropy is also sensitive to $\phi_{P,q}$. When $\cos 2\phi_{P,q}>0$ (see the upper left plot), the entropy seems to decrease overall, but in some regions it actually increases.  On the other hand, when $\cos 2\phi_{P,q}<0$,  the entropy increases towards $\phi_{P,q}\to  \frac{\pi}{2}$. This shows   that the novel kinematical effects due to the intrinsic  transverse momentum of the target, absent in  pure collinear factorization, introduces an additional handle for  generating  quantum resources.

\section{Conclusion and Outlook} 
In this work, we calculated the spin density matrix of a back-to-back quark-antiquark pair in electron-nucleus scattering in the $k_T$-factorization approach. We have shown that by integrating over the azimuthal angle $\phi_{P,q}$ in the back-to-back limit, the density matrix obtained reduces to that of the one-gluon exchange in collinear factorization. This was somewhat surprising, because it means that the  multiple gluon exchange does not modify the density matrix. However, the present calculation is more general and, for differential observables in $\phi_{P,q}$, is sensitive to the linearly polarized WW gluon distribution. We have therefore calculated the density matrix coefficients associated with this second contribution. 

At the (final state) partonic level, we found that massless quark-antiquark pairs (see Fig.~\ref{fig:Concurrence_AngularAve_TL}, $\beta=1$) produced in DIS are always spin-entangled and exhibit Bell-nonlocality. However, for massless quarks the measurement of the spin density matrix cannot rely on weak decays.  
For heavy quarks,  only a small fraction of momentum is lost in the hadronization process. As a consequence, to first approximation, one has $z^{\rm quark} \approx z^{\rm hadron}$ and $k_\perp^{\rm quark} \approx k_\perp^{\rm hadron}$. Moreover, the spin of the (anti)quark is largely transferred to the produced (anti)baryon~\cite{Galanti:2015pqa}. With less strong decorrelation patterns, one can opt for the measurement channels already proposed in the literature Ref.~\cite{Galanti:2015pqa,Qi:2025onf}. 

From quantum informational perspective, the density matrix for massive quarks is much richer and more complex,  depending on multiple kinematical parameters such as $\varepsilon$, $Q^2$ and $q_\perp$. For the first time,  we have confirmed that states with non-vanishing magic are  produced in inclusive production, although the largest value of $S_{\rm R}^{\rm stab}\sim 0.45$ found in this work is somewhat less than in the case of exclusive production  $S_{\rm R}^{\rm stab}\sim 0.58$ \cite{Hatta:2025obw}. We have also found that,  for bottom quark pairs  with  $q_{\perp}\sim$ (a few GeV), the distributions of entanglement and  magic  are anisotropic in the azimuthal angle $\phi_{P,q}$ due to the linearly polarized gluon distribution. At  $\phi_{P,q}=\frac{\pi}{2}$,  we observe the largest amount of entanglement and magic.   

In the future, it will be important to quantify various final-state effects (such as Sudakov logarithms~\cite{Hatta:2020bgy,Hatta:2021jcd,Marquet:2025jdr,Gu:2025ijz}) and also compute higher-order corrections to these observables, to bring phenomenological studies to a precision level.

\section*{Acknowledgments}
The work of M. F. is supported by the ULAM fellowship program of NAWA No. BNI/ULM/2024/1/00065 ``Color glass condensate effective theory beyond the eikonal approximation''. Y.~H. is supported by the U.S. Department
of Energy under Contract No. DE-SC0012704, by LDRD funds from Brookhaven Science Associates, and also by the framework of the Saturated Glue (SURGE) Topical Theory Collaboration. B.X. is supported in part by the Ministry of Science and Technology of China under Grant No. 2024YFA1611004, by the Natural Science Foundation of Guangdong Province under Grant No.~2026A1515011242. 

\appendix

\section{Inclusive dijet production in the Color Glass Condensate}
\label{App:DijetCross}
\subsection{Impact factors}
The $S$-matrix element for the inclusive dijet production in DIS reads
\begin{equation}
    S_{\gamma^{*} \rightarrow q_1  q_2 } = - i e e_f \varepsilon_{\mu}^{\lambda} ( p_{\gamma}) \int d^4 z \; \theta (-z^+) e^{i p_{\gamma} \cdot z} \; \bar{u}^{\rm shock}_{\alpha} (p_q, z) \gamma^{\mu} v_{\alpha'}^{\rm shock} (p_{\bar{q}}, z)
    \label{Eq:DijetSMatrix}
\end{equation}
where the effective quark/anti-quark spinors through the shockwave reads
\begin{gather}
    \bar{u}^{\rm shock}_{\alpha} (p_q, z) = \theta (p_q^+) \int \frac{d^2 k_{\perp}}{(2 \pi)^2} \int d^2 x_{1 \perp} e^{-i x_{1 \perp} \cdot ( k_1 - k)_{\perp} } \mathcal{U}_F ( x_{1 \perp} ) \nonumber \\ \times e^{i z^+ \frac{k_{\perp}^2 + m^2}{2 p_q^+} + i z^- p_q^+ - i z_{\perp} \cdot k_{\perp} } \bar{u}_{\alpha} (p_q) \gamma^+ \frac{ \gamma^- p_q^+ - \Slash{k}_{\perp} + m }{2 p_q^+} \; ,
\end{gather}

\begin{gather}
    v^{\rm shock}_{\alpha'} (p_{\bar{q}}, z) = - \theta (-p_{\bar{q}}^+) \int \frac{d^2 p_{\perp} }{(2 \pi)^2} \int d^2 x_{2 \perp} e^{-i x_{2 \perp} \cdot ( k_2 - p)_{\perp} } \mathcal{U}_F^{\dagger} ( x_{2 \perp} ) \nonumber \\ \times e^{i z^+ \frac{ p_{\perp}^2 + m^2}{2 p_{\bar{q}}^+} + i z^- p_{\bar{q}}^+ - i z_{\perp} \cdot p_{\perp} } \; \frac{ \gamma^- p_{\bar{q}}^+ - \Slash{p}_{\perp} - m }{2 p_{\bar{q}}^+} \gamma^+ v_{\alpha'} (p_{\bar{q}}) \; .
\end{gather}
Integrating over $z^+, z^-$ and $z_{\perp}$ in eq.~(\ref{Eq:DijetSMatrix}), one can easily get
\begin{gather}
   \mathcal{M}_{\gamma^{*} \rightarrow q_1  q_2 } = \frac{(-i) S_{\gamma^{*} \rightarrow q_1  q_2 }}{2 p_{\gamma}^+ (2 \pi) \delta (p_{\gamma}^+ - p_q^+ - p_{\bar{q}}^+ ) } \nonumber \\ = (-i) e e_f \varepsilon_{\mu}^{\lambda} (p_{\gamma}) \int d^2 x_{1 \perp}  \int d^2 x_{2 \perp} e^{-i x_{1 \perp} \cdot p_{q \perp} } e^{-i x_{2 \perp} \cdot p_{\bar{q}}^+ } \Phi^{\mu} (x_{12 \perp}) [\mathcal{U}_F ( x_{1 \perp} ) \mathcal{U}_F^{\dagger} (x_{2 \perp}) - 1] \; ,
   \label{Eq:MDijet}
\end{gather}
where
\begin{gather}
    \Phi^{\mu} (x_{12 \perp}) = \frac{p_q^+ (p_{\gamma}^+ - p_q^+)}{(p_{\gamma}^+)^2} \int \frac{d^2 k_{\perp} }{(2 \pi)^2} \frac{ e^{i k_{\perp}  \cdot x_{12 \perp} } }{ k_{\perp}^2 + z\bar{z}Q^2+m^2 } \nonumber \\ \times \bar{u}_{\alpha} (p_q) \gamma^+ \frac{ \gamma^- p_q^+ - \Slash{k}_{\perp} + m }{2 p_q^+} \gamma^{\mu} \frac{\gamma^- (p_{\gamma}^+ - p_q^+ ) + \Slash{k}_{\perp} - m}{2 (p_{\gamma}^+ - p_q^+)} \gamma^+  v_{\alpha'} (p_{\bar{q}}) \; .
\end{gather}
For a longitudinal photon, we can set
\begin{equation}
    \varepsilon_{\mu}^{\lambda} (q) \rightarrow \varepsilon_{\mu}^{+} (q) \equiv \frac{Q}{q^+} g_{\mu}^+
\end{equation}
and we obtain
\begin{gather}
  \Phi^{L}_{\alpha \alpha'} (x_{12 \perp}) \equiv  \varepsilon_{\mu}^{+} (q) \Phi^{\mu} (x_{12 \perp}) =  \frac{z \bar{z}  Q}{q^+} \int \frac{d^2 k_{\perp} }{(2 \pi)^2} \frac{ e^{i k_{\perp} \cdot x_{12 \perp} } }{ k_{\perp}^2 + z\bar{z}Q^2+m^2 } \bar{u}_{\alpha} (p_q)  \gamma^+  v_{\alpha'} (p_{\bar{q}})  \; ,
\end{gather}
where we defined $z = p_q^+/p_{\gamma}^+$ and $\bar{z} = 1-z$. In the transverse case, we obtain
\begin{gather}
  \Phi^{T(\lambda)}_{\alpha \alpha'} (x_{12 \perp})  \equiv  \varepsilon^i_{\perp \lambda} \Phi^{i} (x_{12 \perp}) = \frac{1}{2 q^+} \int \frac{d^2 k_{\perp} }{(2 \pi)^2} \frac{ e^{i k_{\perp} \cdot x_{12 \perp} } }{ k_{\perp}^2 + z\bar{z}Q^2+m^2 } \nonumber \\ \times \bar{u}_{\alpha} (p_q) \left[ (1-2 z) k_{\perp}^i + i \epsilon^{ij} k_{\perp}^j \gamma^5 - m \gamma_{\perp}^i \right] \gamma^+  v_{\alpha'} ( p_{\bar{q}} ) \; .
\end{gather}
The $\Phi^{T(\lambda)}_{\alpha \alpha'} (x_{12 \perp})$ are Fourier transform, i.e. 
\beq
\Phi^{L/T(\lambda)}_{\alpha \alpha'} (x_{12\perp}) = \int \frac{d^2 k_{\perp}}{(2 \pi)^2} e^{i x_{12 \perp} \cdot k_{\perp}} \Phi^{L/T(\lambda)}_{\alpha \alpha'} (k_{\perp}) \; ,
\eeq
of the momentum space impact factors 
\beq
\Phi^{L}_{\alpha \alpha'} (P_\perp)= \frac{z\bar{z}Q}{p_{\gamma}^+} \frac{1}{P_\perp^2+z\bar{z}Q^2+m^2} \bar{u}_{\alpha} (p_q)\gamma^+ v_{\alpha'} (p_{\bar{q}}) \; ,
\label{Eq:PhiLon}
\eeq
and
\beq
\Phi^{T(\lambda)}_{\alpha \alpha'} (P_\perp) &=&\frac{1}{2p_{\gamma}^+} \frac{\varepsilon^i_\lambda}{P_\perp^2+z\bar{z}Q^2+m^2} \bar{u}_{\alpha} (p_q) \left((1-2z)P_\perp^i+i\epsilon^{ij}P_\perp^j \gamma_5-m\gamma_\perp^i\right) \gamma^+ v_{\alpha'} (p_{\bar{q}}) \: .
\label{Eq:PhiTra}
\eeq

\subsection{Cross section of the inclusive dijet production}
The CGC event-by-event amplitude can be written 
\begin{gather}
   \mathcal{M}_{\gamma^{*} \rightarrow q_1  q_2 }^{L/T(\lambda)} = (-i) e e_f \int d^2 x_{1 \perp}  \int d^2 x_{2 \perp} e^{-i x_{1 \perp} \cdot p_{q \perp} } e^{-i x_{2 \perp} \cdot p_{\bar{q} \perp} } \Phi^{L/T(\lambda)}_{\alpha \alpha'} (x_{12\perp}) [\mathcal{U}_F ( x_{1 \perp}) \mathcal{U}_F^{\dagger} (x_{2 \perp}) - 1] \; ,
\end{gather}
The longitudinal and transverse cross sections are respectively obtained as
\beq
   \frac{d \sigma^{L} }{d P.S. } = 2 p_{\gamma}^+ (2 \pi) \delta (p_{\gamma}^+ - p_q^+ - p_{\bar{q}}^+ )  \sum_{\rm hel. \; col.} \langle |\mathcal{M}_{\gamma^{*} \rightarrow q_1  q_2 }^{L} |^2 \rangle_{A} \; ,
\eeq
\beq
   \frac{d \sigma^{T} }{d P.S. } = 2 p_{\gamma}^+ (2 \pi) \delta (p_{\gamma}^+ - p_q^+ - p_{\bar{q}}^+) \frac{1}{2} \sum_{\rm hel. \; col.} \langle |\mathcal{M}_{\gamma^{*} \rightarrow q_1  q_2 }^{T} |^2 \rangle_{A} \; ,
\eeq
where $d P.S.$ denotes the two-particle Lorentz-invariant phase space, $\langle  ... \rangle_A$ is the CGC averaging, and, in the transverse case, we average over the initial photon polarization $\lambda$. We thus obtain
\begin{gather}
     \frac{d \sigma^{L/T} }{d z d^2 p_{q \perp} d^2 p_{\bar{q} \perp} } =   N_c \alpha_{em} e_f^2 \; p_{\gamma}^+ \int \frac{d^2 x_{1 \perp}}{(2 \pi)^2}  \int \frac{ d^2 x_{2 \perp}}{(2 \pi)^2} \int \frac{d^2 x_{1'  \perp}}{(2 \pi)^2}  \int \frac{d^2 x_{2' \perp}}{(2 \pi)^2} e^{-i p_{q \perp} \cdot x_{1 1' \perp} } e^{-i p_{ \bar{q} \perp} \cdot x_{2 2' \perp} } \nonumber \\  \psi^{L/T (\lambda)}_{\alpha \alpha'} (x_{12\perp}) \psi^{L/T (\lambda) *}_{\beta \beta'} (x_{1'2'\perp})   ( 1 + S^{(4)} (x_{1 \perp}, x_{2 \perp}; x_{1 \perp}, x_{2 \perp}) - S^{(2)} (x_{1 \perp}, x_{2 \perp}) -  S^{(2)} (x_{2' \perp}, x_{1' \perp}) ) \; ,
\end{gather}
where we defined the longitudinal and transverse splitting wave function 
\beq
 \psi^{L}_{\alpha \alpha'} (x_{12\perp})  \equiv (2 \pi)^2  \sqrt{\frac{p_{\gamma}^+}{p_q^+ p_{\bar{q}}^+}}  \Phi^{L}_{\alpha \alpha'} (x_{12\perp}) \; ,
\eeq
\beq
 \psi^{T(\lambda)}_{\alpha \alpha'} (x_{12\perp})  \equiv (2 \pi)^2  \sqrt{\frac{p_{\gamma}^+}{2 p_q^+ p_{\bar{q}}^+}}  \Phi^{T(\lambda)}_{\alpha \alpha'} (x_{12\perp}) \; ,
\eeq
which exactly match the normalization of those of \cite{Dominguez:2011wm}. 

\section{Calculation of the spin-density matrix in the transverse photon case}
\label{App:DetailTransvPhoton}

Let us start by considering
\begin{gather}
\frac{\partial}{\partial  P_{\perp}^j } \psi_{\alpha \alpha'}^{T(\lambda)} (P_\perp) = \sqrt{c_T} \; \varepsilon^k_\lambda \frac{\partial}{\partial P_{\perp }^j }\frac{1}{ P_{\perp}^2 + \bar{Q}^2 } \bar{u}_{\alpha} ( p_q ) \left[ (1-2z) P_{\perp}^k  + i\epsilon^{kl} P_{\perp}^{l} \gamma^5  - m \gamma^k   \right] \gamma^+ v_{\alpha'}( p_{\bar{q}}) \label{delphi} \; , 
\end{gather}
\begin{gather}
\frac{\partial}{\partial  P_{\perp}^i } \psi_{\beta \beta'}^{T(\lambda)*} (P_\perp) =  \sqrt{c_T} \; \varepsilon^k_\lambda \frac{\partial}{\partial P_{\perp}^i}\frac{1}{ P_{\perp}^2+ \bar{Q}^2 } \bar{v}_{\beta'} ( p_{\bar{q}}) \left[ (1-2z) P_{\perp}^k -i \epsilon^{kl} P_{\perp}^{l} \gamma^5 + m  \gamma^k \right] \gamma^+  u_{\beta} ( p_q) \nonumber \; ,
\end{gather}
where, for convenience, we introduced
\beq
   c_{T} =   \frac{(2 \pi)^4 }{8 p_q^+ p_{\bar{q}}^+ p_{\gamma}^+ } \; , \hspace{1.5 cm} \bar{Q} = \sqrt{z \bar{z} Q^2 + m^2} \; .
   \label{Eq:UsefConst}
\eeq
Taking the product and summing over the two possible polarizations, we find
\begin{gather*}
\left[ \frac{\partial}{\partial  P_{\perp}^i} \psi_{\beta\beta'}^{T(\lambda)*} (P_\perp) \right]  \left[ \frac{\partial}{\partial P_{\perp }^j } \psi^{T (\lambda)}_{\alpha\alpha'} (P_\perp) \right]  \nonumber \\ = c_{T} \left[ \frac{\partial}{\partial P_{\perp}^i}\frac{1}{ P_{\perp}^2 + \bar{Q}^2} \bar{v}_{\beta'} ( p_{\bar{q}}) \left[ (1-2z) P_{\perp}^k -i \epsilon^{kl} P_{\perp}^{l} \gamma^5 + m  \gamma^k \right] \gamma^+  u_{\beta} (p_q) \nonumber \right] \nonumber \\ \times \left[ \frac{\partial}{\partial P_{\perp}^j }\frac{1}{ P_{\perp}^2 + \bar{Q}^2} \bar{u}_{\alpha} ( p_q ) \left[ (1-2z) P_{\perp}^k  + i\epsilon^{kn} P_{\perp}^n \gamma^5  - m \gamma^k   \right] \gamma^+ v_{\alpha'}( p_{\bar{q}}) \right] \nonumber \\
    = c_{T} \Bigg \{ \frac{ 4 P_{\perp}^i P_{\perp}^j}{[ P_{\perp}^2 + \bar{Q}^2]^4} \bar{v}_{\beta'} ( p_{\bar{q}}) \left[ (1-2z) P_{\perp}^k -i \epsilon^{kl} P_{\perp}^{l} \gamma^5 + m  \gamma^k \right] \gamma^+  u_{\beta} ( p_q) \nonumber \\ \times \bar{u}_{\alpha} ( p_q ) \left[ (1-2z) P_{\perp}^k  + i\epsilon^{kn} P_{\perp}^n \gamma^5  - m \gamma^k   \right] \gamma^+ v_{\alpha'}( p_{\bar{q}}) \nonumber \\ - \frac{ 2 P_{\perp}^i }{[ P_{\perp}^2 + \bar{Q}^2 ]^3} \bar{v}_{\beta'} ( p_{\bar{q}}) \left[ (1-2z) P_{\perp}^k - i \epsilon^{kl} P_{\perp}^{l} \gamma^5 + m  \gamma^k \right] \gamma^+  u_{\beta} ( p_q)  \bar{u}_{\alpha} ( p_q ) \left[ (1-2z) \delta^{kj}  + i \epsilon^{kj} \gamma^5   \right] \gamma^+ v_{\alpha'}( p_{\bar{q}}) \nonumber \\ - \frac{ 2 P_{\perp}^j }{[ P_{\perp}^2 + \bar{Q}^2 ]^3}  \bar{v}_{\beta'} ( p_{\bar{q}}) \left[ (1-2z) \delta^{ki} -i \epsilon^{ki} \gamma^5 \right] \gamma^+  u_{\beta} ( p_q) \bar{u}_{\alpha} ( p_q ) \left[ (1-2z) P_{\perp}^k  + i\epsilon^{kn} P_{\perp}^n \gamma^5  - m \gamma^k   \right] \gamma^+ v_{\alpha'}( p_{\bar{q}}) \nonumber \\ + \frac{ 1 }{ [P_{\perp}^2 + \bar{Q}^2]^2} \bar{v}_{\beta'} ( p_{\bar{q}}) \left[ (1-2z) \delta^{ki} -i \epsilon^{ki} \gamma^5 \right] \gamma^+  u_{\beta} ( p_q)  \bar{u}_{\alpha} ( p_q ) \left[ (1-2z) \delta^{kj}  + i \epsilon^{kj} \gamma^5   \right] \gamma^+ v_{\alpha'}( p_{\bar{q}}) \bigg \} \; .
\end{gather*}
We now write
\begin{gather}
\left[ \frac{\partial}{\partial  P_{\perp}^i} \psi_{\beta\beta'}^{T(\lambda)*} (P_\perp) \right]  \left[ \frac{\partial}{\partial P_{\perp }^j } \psi^{T (\lambda)}_{\alpha\alpha'} (P_\perp) \right] 
    = c_{T} \left[ F_{\alpha \alpha' \beta \beta'}^{ij} + S_{\alpha \alpha' \beta \beta'}^{ij} + T_{\alpha \alpha' \beta \beta'}^{ij} \right] \; .
\label{Eq:FSTdec}
\end{gather}
The first term on the right-hand side of eq.~(\ref{Eq:FSTdec}) reads
\begin{gather*}
   F_{\alpha \alpha' \beta \beta'}^{ij} = \frac{ 4 P_{\perp}^i P_{\perp}^j}{[P_{\perp}^2 + \bar{Q}^2]^4} \bar{v}_{\beta'} ( p_{\bar{q}}) \left[ (1-2z) P_{\perp}^k -i \epsilon^{kl} P_{\perp}^{  l}\gamma^5 + m  \gamma^k \right] \gamma^+  u_{\beta} ( p_q)  \nonumber \\ \times \bar{u}_{\alpha} ( p_q ) \left[ (1-2z) P_{\perp}^k  + i\epsilon^{kn} P_{\perp}^n \gamma^5  - m \gamma^k   \right] \gamma^+ v_{\alpha'}( p_{\bar{q}}) \nonumber \\ = \frac{ 4 P_{\perp}^i P_{\perp}^j}{[ P_{\perp}^2 + \bar{Q}^2]^4} \Bigg \{ P_{\perp}^2 \bigg( (1 - 2 z)^2  \bar{v}_{\beta'} ( p_{\bar{q}}) \gamma^+  u_{\beta} ( p_q)  \bar{u}_{\alpha} ( p_q ) \gamma^+ v_{\alpha'}( p_{\bar{q}}) + \bar{v}_{\beta'} ( p_{\bar{q}}) \gamma^5  \gamma^+  u_{\beta} ( p_q)  \bar{u}_{\alpha} ( p_q )  \gamma^5  \gamma^+ v_{\alpha'}( p_{\bar{q}})  \bigg ) \nonumber \\ - m^2 \bar{v}_{\beta'} ( p_{\bar{q}})   \gamma^k \gamma^+  u_{\beta} ( p_q)  \bar{u}_{\alpha} ( p_q ) \gamma^k    \gamma^+ v_{\alpha'}( p_{\bar{q}})  \nonumber \\ - (1-2z)  m |P_{\perp}| \bigg( \bar{v}_{\beta'} ( p_{\bar{q}}) \gamma^+  u_{\beta} ( p_q)  \bar{u}_{\alpha} ( p_q )  \gamma^1 \gamma^+ v_{\alpha'}( p_{\bar{q}}) - \bar{v}_{\beta'} ( p_{\bar{q}})  \gamma^1 \gamma^+  u_{\beta} ( p_q)  \bar{u}_{\alpha} ( p_q )   \gamma^+ v_{\alpha'}( p_{\bar{q}}) \bigg)  \nonumber \\  - i m |P_{\perp}| \bigg( \bar{v}_{\beta'} ( p_{\bar{q}}) \gamma^5  \gamma^+  u_{\beta} ( p_q)  \bar{u}_{\alpha} ( p_q ) \gamma^2   \gamma^+ v_{\alpha'}( p_{\bar{q}}) + \bar{v}_{\beta'} ( p_{\bar{q}})  \gamma^2  \gamma^+  u_{\beta} ( p_q)  \bar{u}_{\alpha} ( p_q ) \gamma^5 \gamma^+ v_{\alpha'}( p_{\bar{q}}) \bigg) \Bigg \} \nonumber \\ \equiv \delta_{ij} F_{\alpha \alpha' \beta \beta'} + \left( \frac{ 2 P_{\perp}^i P_{\perp}^j}{P_{\perp}^2} - \delta_{ij} \right) \; F_{\alpha \alpha' \beta \beta'} \; ,
\end{gather*}
where the various coefficients of the spin-density matrix $F_{\alpha \alpha' \beta \beta'}$ can be explicitly evaluated by the procedure in \cite{Fucilla:2025kit} that we also adopted for the longitudinal case. \\  

The second term in eq.~(\ref{Eq:FSTdec}) reads
\begin{gather*}
  S_{\alpha \alpha' \beta \beta'}^{ij} = - \frac{ 2 P_{\perp}^i }{[ P_{\perp}^2 + \bar{Q}^2 ]^3} \bar{v}_{\beta'} ( p_{\bar{q}}) \left[ (1-2z) P_{\perp}^k -i \epsilon^{kl} P_{\perp}^{l} \gamma^5 + m  \gamma^k \right] \gamma^+  u_{\beta} ( p_q)  \bar{u}_{\alpha} ( p_q ) \left[ (1-2z) \delta^{kj}  + i \epsilon^{kj} \gamma^5   \right] \gamma^+ v_{\alpha'}( p_{\bar{q}}) \nonumber \\ - \frac{ 2 P_{\perp}^j }{[ P_{\perp}^2 + \bar{Q}^2 ]^3}  \bar{v}_{\beta'} ( p_{\bar{q}}) \left[ (1-2z) \delta^{ki} -i \epsilon^{ki} \gamma^5 \right] \gamma^+  u_{\beta} ( p_q) \bar{u}_{\alpha} ( p_q ) \left[ (1-2z) P_{\perp}^k  + i\epsilon^{kn} P_{\perp}^n \gamma^5  - m \gamma^k   \right] \gamma^+ v_{\alpha'}( p_{\bar{q}})
\end{gather*}
We can decompose $S_{\alpha \alpha' \beta \beta'}^{ij}$ as 
\begin{equation}
    S_{\alpha \alpha' \beta \beta'}^{ij} = \delta^{ij} S_{1 ; \alpha \alpha' \beta \beta'} + \left(\frac{2 P^i P^j}{ P_{\perp} ^2}-\delta^{ij}\right) S_{2; \alpha \alpha' \beta \beta'} \; ,  
\end{equation}
where
\begin{equation}
    S_2 = \left(\frac{P^k P^l}{ P_{\perp}^2} - \frac{ \delta^{kl} }{2}\right) S^{kl} 
\end{equation}
It is easy to note that $S_{1; \alpha \alpha' \beta \beta' } = S_{2; \alpha \alpha' \beta \beta' } \equiv S_{ \alpha \alpha' \beta \beta' } $, so that 
\begin{equation}
    S_{\alpha \alpha' \beta \beta'}^{ij} = \delta^{ij} S_{ \alpha \alpha' \beta \beta'} + \left(\frac{2 P^i P^j}{ P_{\perp} ^2}-\delta^{ij}\right) S_{\alpha \alpha' \beta \beta'} \; , 
\end{equation}
The spin-density matrix $S_{ \alpha \alpha' \beta \beta'}$ is
\begin{gather*}
    S_{ \alpha \alpha' \beta \beta' } = \frac{\delta^{ij}}{2} S_{\alpha \alpha' \beta \beta' }^{ij} \nonumber \\  = - \frac{ 1  }{[ P_{\perp}^2 + \bar{Q}^2 ]^3} \Bigg \{ 2 P_{\perp}^2 \bigg( (1 - 2 z)^2  \bar{v}_{\beta'} ( p_{\bar{q}}) \gamma^+  u_{\beta} ( p_q)  \bar{u}_{\alpha} ( p_q ) \gamma^+ v_{\alpha'}( p_{\bar{q}}) + \bar{v}_{\beta'} ( p_{\bar{q}}) \gamma^5  \gamma^+  u_{\beta} ( p_q)  \bar{u}_{\alpha} ( p_q )  \gamma^5  \gamma^+ v_{\alpha'}( p_{\bar{q}})  \bigg ) \nonumber \\ - (1-2z)  m | P_{\perp} | \bigg( \bar{v}_{\beta'} ( p_{\bar{q}}) \gamma^+  u_{\beta} ( p_q)  \bar{u}_{\alpha} ( p_q )  \gamma^1 \gamma^+ v_{\alpha'}( p_{\bar{q}}) - \bar{v}_{\beta'} ( p_{\bar{q}})  \gamma^1 \gamma^+  u_{\beta} ( p_q)  \bar{u}_{\alpha} ( p_q )   \gamma^+ v_{\alpha'}( p_{\bar{q}}) \bigg)  \nonumber \\  - i m | P_{\perp} | \bigg( \bar{v}_{\beta'} ( p_{\bar{q}}) \gamma^5  \gamma^+  u_{\beta} ( p_q)  \bar{u}_{\alpha} ( p_q ) \gamma^2   \gamma^+ v_{\alpha'}( p_{\bar{q}}) + \bar{v}_{\beta'} ( p_{\bar{q}})  \gamma^2  \gamma^+  u_{\beta} ( p_q)  \bar{u}_{\alpha} ( p_q ) \gamma^5 \gamma^+ v_{\alpha'}( p_{\bar{q}}) \bigg) \Bigg \} \; .
\end{gather*}
Again, the various coefficients of the spin-density matrix $S_{\alpha \alpha' \beta \beta'}$ can be explicitly evaluated by the procedure in \cite{Fucilla:2025kit}. \\ 

Lastly, the third term in eq.~(\ref{Eq:FSTdec}) reads
\begin{gather*}
  T_{ \alpha \alpha' \beta \beta' }^{ij} =  \frac{ 1 }{[ P_{\perp}^2 + \bar{Q}^2]^2} \bar{v}_{\beta'} ( p_{\bar{q}}) \left[ (1-2z) \delta^{ki} -i \epsilon^{ki} \gamma^5 \right] \gamma^+  u_{\beta} ( p_q)  \bar{u}_{\alpha} ( p_q ) \left[ (1-2z) \delta^{kj}  + i \epsilon^{kj} \gamma^5   \right] \gamma^+ v_{\alpha'}( p_{\bar{q}}) \nonumber \\ = \frac{ \delta^{ij} }{[ P_{\perp}^2 + \bar{Q}^2]^2} \left[ (1-2z)^2    \bar{v}_{\beta'} ( p_{\bar{q}})  \gamma^+  u_{\beta} ( p_q)  \bar{u}_{\alpha} ( p_q )  \gamma^+ v_{\alpha'}( p_{\bar{q}}) + \bar{v}_{\beta'} ( p_{\bar{q}}) \gamma^5 \gamma^+  u_{\beta} ( p_q)  \bar{u}_{\alpha} ( p_q ) \gamma^5 \gamma^+ v_{\alpha'}( p_{\bar{q}}) \right] \nonumber \\ \equiv \delta^{ij} T_{ \alpha \alpha' \beta \beta' } \; .
\end{gather*}
Thus, we obtain
\begin{gather}
    \left[ \frac{\partial}{\partial  P_{\perp}^i} \psi_{\beta\beta'}^{T(\lambda)*} (P_\perp) \right]  \left[ \frac{\partial}{\partial P_{\perp }^j } \psi^{T (\lambda)}_{\alpha\alpha'} (P_\perp) \right] = c_T \bigg \{ \delta^{ij} \; (F_{ \alpha \alpha' \beta \beta' }+S_{ \alpha \alpha' \beta \beta' }+T_{ \alpha \alpha' \beta \beta' }) \nonumber \\ + \left(\frac{2 P^i P^j}{P_{\perp}^2} - \delta^{ij}\right) (F_{ \alpha \alpha' \beta \beta' } + S_{ \alpha \alpha' \beta \beta' }) \bigg \} \; ,
\end{gather}
which, by comparison with
\beq
\left[ \frac{\partial}{\partial  P_{\perp}^i } \psi_{\beta\beta'}^{T(\lambda)*} (P_\perp) \right]  \left[ \frac{\partial}{\partial P_{\perp }^j } \psi^{T(\lambda)}_{\alpha\alpha'} (P_\perp) \right] 
&=&\delta^{ij}\Bigl[A^{(0)} +   \tilde{ C}^{(0)}_{ab} \sigma^a \otimes \sigma^b \Bigr] \nonumber \\
 && \hspace{-0.55 cm} +\left(\frac{2P_\perp^i P_\perp^j}{P_\perp^2}-\delta^{ij}\right)\Bigl[A^{(2)} +   \tilde{C}^{(2)}_{ab} \sigma^a \otimes \sigma^b \Bigr] \; , \nonumber 
\eeq
enables the extraction of all coefficients in section \ref{sec:TransvPhoton}.

\section{Entanglement, Bell-nonlocality and magic}
\label{QuantumInfo}
For the convenience of the reader, here we provide the relations useful for quantifying entanglement, Bell-nonlocality, and magic from the spin-density matrix $\rho$. The standard form of the spin density matrix of a two-qubit system reads 
\begin{equation}
    \rho = \frac{1}{4} ( \mathbb{I} \otimes \mathbb{I} + B_a \; \sigma^a \otimes \mathbb{I} + \bar{B}_b \; \mathbb{I} \otimes \sigma^b + C_{ab} \sigma^a \otimes \sigma^b  ) \; .
\label{Eq:GenSpinDens2Qubit}
\end{equation}
In all the cases we consider, $B=\bar{B}=0$. This will be assumed in the following.  

\textbf{Peres-Horodecki criterion for the Entanglement.} Consider the following two quantities 
\beq
\begin{split}
\Delta_1= \sqrt{(C_{rr} -C_{kk} )^2+(C_{rk} +C_{kr} )^2}-1+C_{nn}, \\
\Delta_2= \sqrt{(C_{rr}+C_{kk})^2+(C_{rk}-C_{kr})^2}-1-C_{nn}. 
\end{split}
\eeq
According to the Peres-Horodecki criterion \cite{Peres:1996dw,Horodecki:1997vt}, if one of $\Delta$'s is nonnegative, the $q\bar{q}$ pair is entangled \cite{Afik:2020onf,Afik:2022kwm}. To quantify the entanglement, we can thus use the concurrence, 
\begin{equation}
    \mathcal{C}[\rho] = \max \{ \Delta_1 / 2 , \Delta_2 / 2, 0 \} \; ,
\label{Eq:Concurrence}
\end{equation}
which ranges between 0 and 1.  $\mathcal{C}[\rho]=0$ corresponds to a separable state, while $\mathcal{C}[\rho]=1$ indicates a maximally entangled state. \\

\textbf{Bell-CHSH inequality.} Another measure of quantum correlation is the violation of the Bell-CHSH inequality \cite{Clauser:1969ny} 
\beq
{\rm Max}_{\{\vec{n}_i\}}  \Bigl| n_1^a C_{ab} (n_2^b+n_4^b) + n_3^aC_{ab}(n_2^b-n_4^b)
\Bigr| \le 2, \label{ch}
\eeq
for unit vectors $|\vec{n}_{i}|=1$ ($i=1,2,3,4$). Given a  correlation matrix $C_{ab}$, this inequality is violated for certain choices of $\vec{n}_i$ if the largest two of the three eigenvalues $\mu_3\le \mu_2\le \mu_1$ of the matrix $C^{\cal T}C$ (the symbol ${\cal T}$ denotes `transpose') satisfy \cite{Horodecki:1995nsk}
\beq
1<\mu_1+\mu_2\le 2.
\eeq
When this is the case, we say that the pair exhibit `Bell-nonlocality.'

\textbf{Magic.} Finally, we turn to a different entanglement-related quantifier known as magic. It has been established that entanglement by itself is not sufficient to ensure a quantum computational advantage \cite{Gottesman:1998hu}. Rather, quantum computers require entangled states that also possess a non-zero amount of magic in order to achieve a performance beyond classical capabilities. This observation has motivated investigations within the high-energy physics community into whether collider processes can generate particle pairs exhibiting such properties \cite{White:2024nuc,Liu:2025qfl,Gargalionis:2026onv}. To characterize magic in our setup, we employ the stabilizer Rényi entropy \cite{Leone:2021rzd,White:2024nuc}, which for the system under consideration is given by

\begin{gather}
    S_{\rm R}^{\rm stab} = - \ln \left(\frac{1 + C_{n n}^4+C_{r r}^4+C_{k k}^4+C_{k r}^4+C_{r k}^4}{1+C_{n n}^2+C_{r r}^2+C_{k k}^2+C_{k r}^2+C_{r k}^2}\right) \; .
    \label{RenyiStabEntrop}
\end{gather}

\bibliography{ref}
\end{document}